\documentclass[11pt,a4paper]{article}
\pdfoutput=1
\usepackage{jcappub}
\usepackage{amsfonts}
\usepackage{wasysym}
\usepackage{amssymb}
\usepackage{epsfig}
\usepackage{textcomp}
\usepackage{graphicx,psfrag}
\usepackage[caption=false]{subfig}
\usepackage{braket}
\usepackage{color}
\usepackage{placeins}
\usepackage{float}
\usepackage{subfig}
\usepackage{bbold}
\usepackage{booktabs}
\usepackage{longtable}

\title{Observational consistency and future predictions for a 3.5 keV ALP to photon line }
\author[]{Pedro D. Alvarez,}
\author[]{Joseph P. Conlon,}
\author[]{Francesca V. Day,}
\author[]{M.C.~David Marsh,}
\author[]{Markus Rummel}
\affiliation[]{Rudolf Peierls Centre for Theoretical Physics, University of Oxford,\\
1 Keble Road, Oxford, OX1 3NP, United Kingdom}
\emailAdd{alvarez@physics.ox.ac.uk}
\emailAdd{j.conlon1@physics.ox.ac.uk}
\emailAdd{francesca.day@physics.ox.ac.uk}
\emailAdd{david.marsh1@physics.ox.ac.uk}
\emailAdd{markus.rummel@physics.ox.ac.uk}

\abstract{Motivated by the possibility of explaining the 3.5 keV line through dark matter decaying to axion-like particles that subsequently convert to photons, we
study ALP-photon conversion for sightlines passing within 50 pc of the galactic centre. Conversion depends on the galactic centre magnetic field which is highly uncertain. For fields at low or mid-range of observational estimates (10--100 $\mu$G), no observable signal is possible. For fields at the high range of observational estimates (a pervasive poloidal mG field over the central 150 pc) it is possible to generate sufficient signal to explain recent observations of a 3.5 keV line in the galactic centre.
In this scenario, the galactic centre line signal comes predominantly from the region with $ z > 20 \, {\rm pc}$, reconciling the results from the Chandra and XMM-Newton X-ray telescopes. The dark matter to ALP to photon scenario also naturally predicts the non-observation of the 3.5 keV line in stacked galaxy spectra. We further explore predictions for the line flux in galaxies and suggest a set of galaxies that is optimised for observing the 3.5 keV line in this model.

}

\keywords{dark matter, axion, axion-like particle}
\arxivnumber{}

\newcommand{\be}{\begin{equation}}
\newcommand{\ee}{\end{equation}}
\newcommand{\bea}{\begin{eqnarray}}
\newcommand{\eea}{\end{eqnarray}}
\newcommand{\mbb}{\mathbb}
\newcommand{\ti}{\times}

\newcommand{\mc}{\mathcal}

 \newlength{\wth}
\newcommand{\XMM}{{XMM-Newton}~}
\newcommand{\XMMs}{{XMM-Newton}}
\newcommand{\ch}{{Chandra}~}

\newcommand{\dd}{\hbox{d}}

\newcommand{\scen}{DM$\to a\to \gamma$}

\begin{document}

\maketitle
\flushbottom

\section {Introduction}
\label{sec:intro}
Determining the nature and properties of dark matter is one of the most significant challenges of contemporary high-energy physics. Among the variety of strategies for detecting signs of dark matter is the search for unidentified emission lines in the spectra of dark matter dominated astrophysical objects. Such lines could arise in the two-body decay/annihilation of dark matter particles in which the final state contains a photon. Hence, it is not surprising that the recent detection of an unidentified emission line at 3.5 keV in the X-ray spectrum of galaxy clusters and the Andromeda galaxy has generated much interest.

The 3.5 keV line was first observed by two distinct groups, using the detectors of two independent satellites, XMM-Newton and Chandra. After carefully subtracting the astrophysical background of
a stacked sample of galaxy clusters,
reference \cite{Bulbul} found an additional emission line at $E \sim 3.55$ keV, with no apparent astrophysical origin.  The line was observed with both the MOS and PN cameras on the XMM-Newton instrument, and also
reconfirmed in the Perseus cluster using both ACIS-I and ACIS-S configurations on the Chandra satellite.
In \cite{Boyarsky}, a line at very similar energies was also observed in the outskirts of the
Perseus cluster (using a distinct set of XMM-Newton observations than in \cite{Bulbul}) and also in the central region of the Andromeda galaxy (M31).

These papers have triggered many subsequent observational searches for further $E \sim 3.5$ keV line emission.
In \cite{RiemerSorensen}, a search was performed in the spectrum of Chandra observations of the central region of the Milky Way, finding no evidence
of such line emission. More recently, two further analyses of
the galactic centre spectrum using XMM-Newton data have been carried out  \cite{JP, BoyarskyII}. These papers find mutually consistent results, but differ significantly in interpretations. Both analyses find
evidence for a line around 3.5 keV.
The authors of reference
\cite{JP} argue that this line should be attributed to K XVIII emission at 3.47 and 3.51 keV, while the authors of reference \cite{BoyarskyII}
find that the line is consistent with
 a dark matter interpretation, while agreeing that the complicated galactic centre environment
does not allow for definitive exclusion of astrophysical interpretations. In both analyses,
the line flux per arcmin$^{2}$ observed with XMM-Newton is above the upper bound set by non-observations with Chandra in \cite{RiemerSorensen}.

Furthermore, reference \cite{JP} also challenged the existence of a line in M31 (claimed in \cite{Boyarsky}), finding only 1$\sigma$ support for such a line, and questioned
the existence of an unexplained line from clusters, arguing that by including K XVIII lines at 3.47 and 3.51 keV  and a Cl XVII line at 3.51 keV,  no significant excess around 3.5 keV could be established after
allowing for systematic uncertainties.
In response, the authors of reference \cite{Boyarsky} pointed out in \cite{BoyarskyIII} that
the lower significance of the 3.5 keV line for the M31 analysis in \cite{JP} was due to a restriction to an
inappropriately narrow fitting interval of 3--4 keV, resulting in a relatively poorer
fit for the index of the power law background and an inevitably lower significance for the line signal.
In reference \cite{Bulbul2}, the authors of \cite{Bulbul} robustly disagreed with the analysis of cluster data in \cite{JP}, arguing that they
 relied upon incorrect atomic data and inconsistent spectroscopic modelling.

Less controversial, yet
equally important, are the null results reported from searches for  the line in galaxies other than M31 and the Milky Way.
In \cite{Malyshev}, a stacked sample of dwarf spheroidal galaxies were observed using XMM-Newton data. Dwarf galaxies are classic dark matter targets as they have high dark matter densities and low background light. Under the sterile neutrino interpretation of the results of \cite{Bulbul}, a signal of the strength reported in \cite{Bulbul}
 was ruled out at a level of 4.6$\sigma$ (or 3.3$\sigma$ under the most conservative assumptions about the galactic dark matter column density). However, no line signal was observed.
Similarly, in
 \cite{Anderson}, a search for the 3.5 keV line was performed in large stacks of archived
 Chandra and XMM-Newton data of galaxies. No evidence of a line at 3.5 keV was found,
and the sterile neutrino interpretation of the line suggested in \cite{Bulbul} was found to be ruled out at 4.4$\sigma$ and 11.8$\sigma$
for Chandra and XMM-Newton samples respectively.

As these same limits apply directly to any model in which dark matter decays directly to the 3.5 keV photon,
these results appear fatal to all models consisting of dark matter decaying directly to photons.  The challenge for
any  dark matter interpretation of the data in \cite{Bulbul} and \cite{Boyarsky}
is to explain why,
\begin{enumerate}
\item
A signal is produced in galaxy clusters, but is absent in the spectrum of  dwarf spheroidals and stacked galaxies.
\item
A signal is observed in the spectrum of M31, but not in other galaxies.
\end{enumerate}

Our focus here is on the scenario proposed in \cite{14032370} -- which we elaborate on below -- which can explain both these features.
In this scenario, dark matter with mass $\sim 7.1$ keV has a predominant decay channel to very light axion-like particles (ALPs), thus giving rise to a 3.5 keV ALP line.  By itself this line is `invisible', but as ALPs convert into photons in astrophysical magnetic fields, a photon line can be generated in regions with relatively large and coherent astrophysical magnetic fields.
The magnetic fields in galaxy clusters are present over large scales ($\mc{O}(1)$ Mpc) and support substantial conversion probabilities, while regular galaxies and dwarf spheroidals give rise to much weaker photon lines as their magnetic fields are only present on $\mc{O}(10)$ kpc scales.

This model was first proposed in \cite{14032370} to explain the morphology of the signal observed in galaxy clusters in \cite{Bulbul}, which is itself in tension with direct dark matter decay to photons. The signal found in \cite{Bulbul} is stronger in the Perseus cluster -- by up to a factor of eight -- than that inferred from the full 73 cluster sample.
The signal in Perseus also comes disproportionately from the central cool core region of the cluster. The signal from the sample of local bright clusters -- Coma, Ophiuchus and Centaurus -- is also notably stronger than for the full stacked sample. As the central regions of cool core galaxy clusters host particularly high magnetic fields, an enhanced signal from central
 cool core regions is a natural prediction of the \scen~model.

Among galaxies, central observations of edge-on spiral galaxies represent the most attractive targets, as the regular disk magnetic field
is both coherent on the scale of the galaxy and, for central observations, orthogonal to the line of sight.
Observational studies \cite{FletcherBeck} suggest that M31 is particularly attractive among edge-on spiral galaxies, as the regular magnetic field is both unusually large (with $B_{\rm reg} \sim B_{\rm random}$) and unusually coherent (with no evidence of reversals among spiral arms). The above puzzling features are therefore consistent with -- and indeed required by -- this model.

Following its introduction in \cite{14032370}, this scenario has been analysed in more detail for the case of the Milky Way halo (excluding the central region) in \cite{14047741} and for the case of cool-core and non-cool-core clusters in \cite{14065518}.
The objectives of the paper are to extend the study of this scenario to the observationally interesting region of the Milky Way centre and to further clarify
its predictions for observations of other galaxies.

This paper is organised as follows.
We start by reviewing the dark matter to ALP to photon (${\rm DM} \to a \to \gamma$) scenario of \cite{14032370} in Section~\ref{scen_section}. In Section~\ref{MWcentre_sec} we review X-ray observations of the galactic centre and review models for the magnetic fields and free electron density in this region. In Section~\ref{sec:results} we study the phenomenology of ALP-photon conversion in the galactic centre. In Section~\ref{sec:external} we consider observations of other galaxies and provide a list of galaxies optimised for the  ${\rm DM} \to a \to \gamma$ scenario and that have been observed by
either Chandra or XMM-Newton.

\section{The dark matter $\to$ ALP $\to$ photon scenario}\label{scen_section}
In this section, we review the scenario proposed in \cite{14032370} in which the 3.5 keV line is produced by dark matter particles decaying into an ALP, $a$, which subsequently converts into X-ray photons in the presence of astrophysical magnetic fields.

\subsection{DM $\to a$}
While the 3.5 keV line -- if indeed caused by decaying dark matter -- may determine the mass of the dark matter particle, it does not by itself
 provide information on its spin and relevant decay channels.
For example,
if the dark matter particle is a fermion, $\psi$, it may decay into a photon and a neutrino through the effective operator $\frac{1}{\Lambda} \psi^{\dagger} \sigma_{\mu \nu} \nu F^{\mu \nu}$.
A frequently  considered case in this category is for $\psi$ to be a right-handed sterile neutrino with a Majorana mass $m_{\psi}$ and a mixing with the  Standard Model neutrinos induced by an operator $yH^{\dagger} \bar L\psi$.

If such a sterile neutrino constitutes dark matter with $m_{\psi} = 7.1$ keV, it would predominantly decay through the `invisible' $Z$-mediated channel $\psi \to \nu \bar \nu \nu$. At the one-loop level, charged currents induce the decay channel $\psi \to \nu \gamma$ with the decay rate,
\be
\Gamma_{\psi \to \nu \gamma} = \frac{9 \alpha_{EM} }{1024 \pi^4} \sin^2(2\theta) G_F^2 m_{\psi}^5 \, , \label{eq:sterile}
\ee
where $\sin \theta = \frac{y v}{ \sqrt{2} m_{\psi}}$ for the Higgs VEV $v/\sqrt{2}$. The total observable photon flux in the field of view is then given by
\be
{\cal F}_{\psi \to \nu \gamma}= \frac{\Gamma_{\psi \to \nu \gamma}}{4 \pi} \int _\text{FOV}  \varrho \, {\rm d} \varrho \, {\rm d} \phi \int_{\rm l.o.s.} \frac{\rho_\text{DM}(l, \varrho, \phi)}{m_{\psi}} {\rm d}l ,
\ee
where $l$ is the distance along the line of sight and $\left( \varrho , \phi \right)$ are cylindrical coordinates within the field of view, with $\varrho$ the angular radial coordinate.

Reference \cite{Bulbul} showed that,  if the unidentified  3.5 keV line  is interpreted as arising from decaying sterile neutrinos, then the mixing angle is determined to be very small but non-vanishing, $\sin^2(2 \theta) \approx 7 \ti 10^{-11}$.
Since then the 3.5 keV signal has also been interpreted in a variety of models in which dark matter decays, annihilates or de-excites with the prompt emission of a photon.
Among these, models involving axion or ALPs as the dark matter particle have been considered in \cite{Higaki,Jaeckel,Lee,Kawasaki,Henning,Higaki2}.

The scenario of \cite{14032370} considered in this paper is however crucially different from these models in that the photon line is a secondary, environmental effect due to the existence of an otherwise `invisible' ALP line. Decay modes for dark matter particles into ALPs exist for both fermionic and scalar dark matter~\cite{14032370}. For example, fermionic dark matter can decay to an ALP and a neutrino through the operator $\frac{\partial_{\mu}a }{\Lambda} \bar \psi \gamma^{\mu} \gamma_5 \nu$ with the decay rate,
\be
\Gamma_{\psi \to \nu a} = \frac{1}{16 \pi} \frac{m_{\psi}^3}{\Lambda^2} \, ,
\ee
which is in principle independent of $\Gamma_{\psi \to \nu \gamma}$ if $\Lambda$ is independent of the mixing angle $\theta$.

In the presence of magnetic fields, ALPs may convert into photons in a process akin to that of neutrino oscillations. As we will now discuss in more detail, in this way a 3.5 keV ALP line may produce an associated 3.5 keV photon line.

\subsection{ $a \to \gamma$}
\label{sec:atogamma}
The relevant interaction term
for axion-photon conversion
 is the Lagrangian operator,
\be
	\mathcal{L}\supset \frac{1}{8 M}a F_{\mu\nu}\tilde{F}^{\mu\nu} \equiv \frac{1}{M}a\vec{E}\cdot\vec{B} \, \equiv g_{a\gamma\gamma} a \vec{E} \cdot \vec{B} \, , \label{eq:op}
\ee
where $g_{a\gamma\gamma} = M^{-1}$ is the ALP photon coupling.
The linearised equations of motion  for a mode of energy $\omega$ propagating in the $x$-direction in the presence of a classical background magnetic field, $\vec B$, are given by,
\cite{Raffelt}
\be
\label{eq:EqofMotion}
\left(\omega  \mbb{1} + \left(\begin{array}{c c c}
			\Delta_{\gamma} & \Delta_{\rm F} & \Delta_{\gamma a y} \\
			\Delta_{\rm F} & \Delta_{\gamma} & \Delta_{\gamma a z} \\
			\Delta_{\gamma a y} & \Delta_{\gamma a z} & \Delta_{a}
		   \end{array}\right) - i\partial_x\right)\left(\begin{array}{c}
								\gamma_y \\
								\gamma_z \\
								a
							      \end{array}\right)= 0 \, .
\ee
Here $\Delta_{\rm F}$ denotes the interaction which induces Faraday rotation between photon polarisation states in an external magnetic field. As we will be concerned with the total photon flux from ALP-photon conversion, we will neglect these terms in the subsequent analysis.

  The  refractive index for photons in a plasma is given by  $\Delta_{\gamma}  = -\omega_{\rm pl}^2/2\omega$, where $\omega_{\rm pl}  = (4\pi\alpha n_e/m_e)^{1/2}$
  is the plasma frequency with $n_e$ the free electron density. The axion-photon mixing induced by the Lagrangian operator of equation \eqref{eq:op} is determined by the matrix elements $\Delta_{\gamma a i}  = B_i/2M$, where $B_i$ denotes the magnetic field in the directions perpendicular to the ALP direction of travel. Finally,  $\Delta_a  = -m_a^2/\omega$ (in this work we assume vanishing ALP mass $m_{a} = 0$).
Formally, we may write the general solution to equation (\ref{eq:EqofMotion}) for an initial state,
 $
  \ket{i} = \left(
  \gamma_y,
  \gamma_z,
 a
 \right)^{\text{T}}\Big|_{x=-L/2}
$
  propagating from $x=-L/2$ to $x=L/2$ as
\be
\label{eq:soln}
\Ket{f} =  {\cal T}_x\left[ \exp \left(-i \omega L \mathbb{1} \ -i \int_{-L/2}^{L/2} \mathcal{M}(x) \dd x\right) \right] 
		     \Ket{i}
		      \, ,
\ee
where
\be
\mathcal{M} (x) = \left(\begin{array}{c c c}
							\Delta_{\gamma}(x) & 0 & \Delta_{\gamma a y}(x) \\
							0 & \Delta_{\gamma}(x) & \Delta_{\gamma a z}(x) \\
							\Delta_{\gamma a y}(x) & \Delta_{\gamma a z}(x) & \Delta_{a}(x)
		   				\end{array}\right) \, .
\ee
Here, ${\mathcal T}_x$ denotes the `$x$-ordering' operator.
For an initially pure ALP state, the ALP-photon conversion is then given by,
\be
P_{a\to \gamma} = |\langle 1,0,0 | f \rangle|^2 +|\langle 0,1,0 | f \rangle|^2 = \left( |\gamma_y|^2 + |\gamma_z|^2 \right) \Big|_{x=L/2} \, .
\ee
In this scenario, the strength of the photon line then depends on both the magnitude and coherence of the magnetic field, and the dark matter column density along the line of sight. The total predicted photon flux is then given by:
\begin{align}
 \begin{aligned}
{\cal F}_{\psi \to \nu \gamma}= \frac{\Gamma_{\text{DM} \to a}}{4 \pi} \int _{\rm FOV} \varrho \, {\rm d} \varrho \, {\rm d} \phi \int_{\rm l.o.s.} \frac{\rho_\text{DM}(l, \varrho, \phi)}{m_\text{DM}} P_{a \to \gamma} \left(l, \varrho, \phi \right) {\rm d}l ,
 \end{aligned}
\end{align}
As we will discuss in Section \ref{sec:results}, ALP-photon conversion in the the Milky Way only proceeds efficiently in the very central region close to Sgr A*, if at all, and the observable photon line flux is then well-approximated by,
\be
  {\cal F}_{DM \to a \to \gamma}   \simeq \frac{\Omega_{\rm FOV}}{4 \pi \tau_{\rm DM}} \langle P_{a \rightarrow \gamma} \rangle_{\rm  FOV} \int_{l_\text{GC}}^{\infty} \frac{\rho_{\rm DM} \left( l \right) }{m_{\rm DM}} dl \, ,\label{FluxGCgen}
\ee
 where $\langle P_{a\to \gamma} \rangle_{\rm FOV}$ denotes the average conversion probability over the telescope field of view, $\Omega_{\rm FOV}$ is the angular size of the field of view in steradians, and the dark matter density is averaged over the field of view.

\subsection{Predictions of the \scen~scenario}

Here we briefly review predictions made in previous work for the \scen~scenario. With regards to galaxy clusters, reference \cite{14032370} showed that the would-be dark matter decay time assuming DM $\rightarrow$ photons would vary from cluster to cluster in a \scen~scenario, with shorter decay times being inferred for clusters with stronger or more coherent magnetic fields. Within a cluster, the line strength should approximately trace out the square of the magnetic field strength, in particularly peaking strongly in the centre of cool core clusters. The central region of cool core clusters will also give a stronger signal than the central region of non-cool core clusters. These predictions were further discussed and quantified in \cite{14065518}. Reference \cite{14065518} also noted that, due to the increase in magnetic field strength at the centre of a cluster, the would-be decay time inferred from local clusters that fill the field of view will be greater than the decay time inferred from more
distant clusters, where the entire cluster fits in the field of view.

With regards to galaxies, reference \cite{14047741} found that the conversion probability in the Milky Way halo is too low to produce an observable signal, while the conversion probability in M31 is much higher, with M31 displaying highly beneficial conditions for $a \to \gamma$ conversion. Reference \cite{14047741} also predicted a sharp decrease in the signal strength as we move away from the centre of M31, as the magnetic field (following the spiral arms) becomes parallel to the line of sight. Furthermore, \cite{14047741} predicts that the 3.5 keV line signal in a typical galaxy is much weaker than in a galaxy cluster. Reference \cite{14032370} predicted that, among galaxies, edge-on spirals will give the strongest line signal, as the ALP will propagate a larger distance through the disk.

\section{The Milky Way centre}\label{MWcentre_sec}
Motivated by recent analyses of X-ray line emission from the Milky Way centre in  \cite{RiemerSorensen, JP, BoyarskyII}, we here determine
the circumstances under which a signal from the Milky Way centre is achievable in the \scen~model.
ALP to photon conversion in the bulk of the Milky Way has been studied in  \cite{14047741} and found to be too inefficient to contribute significantly to the photon flux.
However, before discussing the predictions of this model, we first review in Section \ref{sec:searches}
the important aspects of the observations \cite{RiemerSorensen, JP, BoyarskyII} in some detail. In \ref{sec:electrondensity} and \ref{sec:magneticfield}, we
review the pertinent aspects of the observational models for the electron density and magnetic field in the central region of the Milky Way.

\subsection{Observational searches for the 3.5 keV line from the galactic centre}\label{sec:searches}

The dynamic centre of the Milky Way is the supermassive black hole associated to the radio source Sgr A*. We take a distance of 8.5 kpc to the galactic centre, and
so $1'$ corresponds to 2.47 pc at the galactic centre.

As it plays an important subsequent role, we first review details of the XMM-Newton and \ch telescopes.
The archival Chandra observations analysed in \cite{RiemerSorensen} involve data from
 ACIS-I configuration which has a
square field of view of $16.8'$ by $16.8'$, consisting of 4 CCD chips I0--I3.
The archival XMM-Newton observations analysed in \cite{BoyarskyII, JP} are with either the XMM-MOS or XMM-PN cameras. These involve
slightly different geometric arrangements of the chips, but in both cases result in a field of view with approximate radius of $15'$.
Observations with MOS1 after 2005 (2012) have reduced coverage due to the failure of one (two) CCDs following micrometeorite damage.

In the \ch observations of \cite{RiemerSorensen},
a 95\% bound on line emission
at $E \sim 3.55~{\rm keV}$ was derived as ${\mc F} \lesssim 3 \ti 10^{-14}~{\rm erg} \, {\rm cm}^{-2} \, {\rm s}^{-1}$, which equates to an upper limit of
${\mc F} \lesssim 5 \ti 10^{-6} \, {\rm photons} \, {\rm cm}^{-2} \, {\rm s}^{-1}$.
The baseline fitted background model also included atomic lines from K XVIII at
 3.48 keV, 3.52 keV and an Ar line at 3.62 keV, and the upper bound on the line flux is sensitive to the
 strengths assigned to these lines. In Table \ref{RS}, we collect the line strengths for the base fit to the data.\footnote{We thank Signe Riemer-S\o rensen
for communicating these to us.}

As the uncertainty in the background modelling is large,
it is possible that the assigned  line strengths may
hide an
actual dark matter signal. A caution on these line strengths is that as they do not come with error bars (due to difficulties of making XSPEC converge) it is possible that there is
actually no statistically significant line emission at these frequencies.
For subsequent comparison with XMM-Newton observations, we re-express these in terms of flux per arcminute$^2$.
In \cite{RiemerSorensen}, the
central $2.5'$  radius around Sgr A* is masked. Hence, for the analysis in Section \ref{sec:results}, we use
an effective field of view of $240$ arcminute$^2$.\footnote{There is further reduction in field of view due to masking of point sources, corresponding to an additional $7\%$ reduction \cite{RiemerSorensen}. We omit this here as a similar point source masking was carried out for XMM-Newton, and we do not know the percentage of field of view lost there.
Given the other uncertainties, this error is minor.}
\begin{table}
\begin{center}
\begin{tabular}{|c|c|c|c|}
\hline
{\rm Element} & {\rm Energy} & {\rm Strength} & {\rm Strength per arcmin}$^{2}$\\
 & ({\rm keV}) & $({\rm ph} \, {\rm cm}^{-2} \, {\rm s}^{-1})$ & $({\rm ph} \, {\rm arcmin}^{-2} \, {\rm cm}^{-2} \, {\rm s}^{-1})$\\
\hline
95 \% Upper bound & 3.55 keV & $\lesssim 5 \ti 10^{-6} $ &  $\lesssim 2.1 \ti 10^{-8}$\\
{\rm K XVIII} & 3.48 & $2.2 \ti 10^{-6}$ & $ 9.2 \ti 10^{-9}$\\
{\rm K XVIII} & 3.52 & $4.2 \ti 10^{-6}$  &  $ 1.8 \ti 10^{-8}$\\
{\rm Ar XVII} & 3.62 & $4.2 \ti 10^{-6}$  &  $ 1.8 \ti 10^{-8}$\\
\hline
\end{tabular}
\end{center}
\caption{Observations of the galactic centre region with Chandra~\cite{RiemerSorensen}. We give the 95 \% upper bound on line emission and also fitted values for atomic
lines included in XSPEC~\cite{xspec} (note that these fitted values are not necessarily statistically distinct from zero)}
\label{RS}
\end{table}

Using archival XMM-Newton data,
references \cite{JP} and  \cite{BoyarskyII} both
detect an emission line at $E \sim 3.5\, {\rm keV}$ with high significance. The former paper however focuses on interpreting
this line in terms of K XVIII emission while the latter paper focuses on a possible dark matter interpretation.
The fluxes as observed by XMM-Newton are shown in Table \ref{JPB}.
We have treated the effective field of view of the MOS and PN chips as 530 arcminute$^2$.
For the MOS camera, this comes from averaging the field of view of MOS1 and MOS2 from the tables in the appendix of~\cite{BoyarskyII}, and we have
assumed the same field of view for the PN camera.
\begin{table}
\begin{center}
\begin{tabular}{|c|c|c|c|}
\hline
Detector & Energy & Strength & Strength per arcmin$^{2}$ \\
 & ({\rm keV}) & $({\rm ph} \, {\rm cm}^{-2} \, {\rm s}^{-1})$ & $({\rm ph} \, {\rm arcmin}^{-2} \, {\rm cm}^{-2} \, {\rm s}^{-1})$\\
\hline
XMM MOS~\cite{JP} & 3.5 & $4.1 \ti 10^{-5} $  &  $ 7.7 \ti 10^{-8} $\\
XMM PN~\cite{JP} & 3.5 & $2.8 \ti 10^{-5} $  &  $ 5.3 \ti 10^{-8} $\\
XMM~\cite{BoyarskyII} & 3.53 & $(2.9 \pm 0.5) \ti 10^{-5} $  &  $ \left( 5.5 \pm 0.9 \right) \ti 10^{-8}$\\
\hline
\end{tabular}
\end{center}
\caption{XMM-Newton observations of the galactic centre region: line emission detected around 3.5 keV}
\label{JPB}
\end{table}

The line strength observed with XMM-Newton is at a level markedly stronger than the upper bound from
Chandra observations. In terms of interpretations involving K XVIII lines, it is unclear
what importance to place on this: the galactic centre environment is complex and multiphase, and it is
conceivable that the regions enclosed by the XMM-Newton field of view
involve a higher average K abundance than those within the Chandra field of view.
However, this would be surprising for the case of dark matter decaying to produce photons.

One aim of this paper is to explain how, in the context of the ${\rm DM} \to a \to \gamma$ scenario, this difference can arise naturally.
In this scenario, the signal is
suppressed within the galactic plane, and so the XMM-Newton field of view, which extends further vertically out of the plane,
contains more signal region. To understand this we now discuss the astrophysics of the galactic centre.

\subsection{Electron density}\label{sec:electrondensity}

As discussed in Section \ref{sec:atogamma}, ALP to photon conversion depends on the free electron density, with large
electron densities suppressing the conversion amplitude. The electron density in the Milky Way centre is therefore an important input into the
resulting signal for the \scen~scenario.

We first describe the coordinates used. We use right-handed Cartesian $(x, y, z)$ coordinates, where the origin $(0, 0, 0)$ corresponds to the centre of galactic coordinates
$(r, b, l) = (8.5\, {\rm kpc}, 0, 0)$. The $x$-coordinate points from the Milky Way centre towards the sun, $y$ is in direction of decreasing $l$ and $z$ points vertically upwards out of the
galactic plane (towards positive $b$). However note that, in these coordinates, the true dynamic marker of the Milky Way centre Sgr A* (where the majority of
observations considered here are centred) is slightly offset, with a physical location of $(l,b)= (-0.06, -0.05)$ \cite{NE2001B}.
This corresponds to $(y_\text{SgrA*}, z_\text{SgrA*}) = (8.9\, {\rm pc}, -7.4\, {\rm pc})$.

In this paper, we will use the NE2001 model for the Milky Way electron density \cite{NE2001}.
 The NE2001 model contains several components,
 and in particular
a galactic centre component that is given in our notation by,
\be
n_{e,\text{GC}}(x,y,z) = 10 \, {\rm cm}^{-3} \exp \left[ - \frac{x^2 + (y - y_\text{GC})^2}{L_\text{GC}^2} \right] \exp \left[ - \frac{(z - z_\text{GC})^2}{H_\text{GC}^2} \right] \, ,
\label{ne2001gc}
\ee
with $L_\text{GC} = 145\, {\rm pc}$ and $H_\text{GC} = 26\, {\rm pc}$. This dominates over thin and thick disk components in the innermost galaxy.
The centroid of the distribution is offset by $y_\text{GC} = 10\, {\rm pc}$ and $z_\text{GC} = -20\, {\rm pc}$. However, note that the physical offset from Sgr A* is reduced
as Sgr A* is itself offset from $x=y=z=0$. Also note in the NE2001 model, the electron density in (\ref{ne2001gc}) is formally truncated to zero when the argument of the exponential is less than -1. However, this truncation reflects an abrupt change in scattering diameters for OH masers in the galactic center, and can be omitted if we are interested only in the
free electron density rather than its fluctuations (see the discussion in Section 2.4 of \cite{Ferriere2007}). We shall therefore use (\ref{ne2001gc}) as our baseline
electron density model in this paper. We also include the thick disk component of \cite{NE2001}, which becomes comparable to the galactic centre component at the edge of our region of interest.

Let us enumerate the caveats on the above electron density.\footnote{More detailed studies of gas distributions within the inner $10\, $pc appear in \cite{Ferriere2012}.}  This electron density is derived via pulsar dispersion and emission measures,
which are sensitive to integrated electron densities along the line of sight. The electron density thus determined
is a smooth function, and does not account for patchiness, or the presence of dense clouds with partial filling factors interspersed by voids.
It is also a single simple function that will represent a fit to data for all lines of sight within $\mc{O}(100)\, {\rm pc}$ from Sgr A*,
while our interest is only in lines of sight enclosed by the fields of view of XMM-Newton and Chandra (extending to a maximum of 37 pc from Sgr A*),
and in particular the regions along them with large transverse magnetic fields.
For all these caveats, the distribution in \cite{NE2001} is nonetheless observationally derived and captures genuine features of the free electron
distribution in the galactic centre. While aware of its limitations, we shall therefore use it in our subsequent studies.

\subsection{Magnetic field}\label{sec:magneticfield}

The magnitude, direction and coherence of the transverse magnetic field in the galactic centre region are clearly important for us to determine the $a\to\gamma$ conversion probability.
Unfortunately, the
magnetic field in the central 100--200 pc of the Milky Way is poorly known, and estimates vary by two orders of magnitude.
As the galactic centre magnetic field in this region arises from different processes to the bulk of the galaxy, this area is excluded from the Milky Way magnetic field model of \cite{Farrar1, Farrar2}.  Following \cite{Davidson1996,Morris2007,Ferriere2009,Ferriere2010, Morris}, we here provide a brief summary of the observational possibilities for the galactic centre magnetic field, from high to low values.

There exists a longstanding case that the magnetic field within the galactic centre is dominantly poloidal (vertical) and with a uniform milligauss strength throughout the central $\sim$150 parsecs \cite{YusefZadeh,YusefZadeh2,Morris1990, Morris1996}. This argument arises from the presence of nonthermal radio filaments in the galactic centre region, orientated predominantly orthogonal to the galactic plane and emitting via synchrotron radiation. These filaments are remarkably straight and uniform, even though some are clearly interacting with molecular clouds. The apparent rigidity of the filaments against collisions with molecular clouds can be used to infer their magnetic pressure, leading to estimates of a $B \sim 1$ mG field strength. This leads to a picture of a pervasive milligauss field, locally illuminated by injection of relativistic electrons.

In \cite{Crocker2010}, the spectrum of synchrotron emission from the galactic centre was used to obtain a minimal magnetic field of $B \sim 50\, \mu$G averaged over a 400 pc region, with a preferred average field of $B \sim 100\, \mu$G over the entire galactic centre region.

In contrast, a much lower estimate is suggested in reference \cite{Ferriere2009}, which argues
for a relatively weak pervasive poloidal magnetic field ($B \sim 10\, \mu$G) within the galactic centre region.
These lower magnetic field estimates are based on equipartition arguments \cite{LaRosa-etal2005}
and on studies of short pc-scale nonthermal radio filaments \cite{YusefZadeh-etal2004}.
On this view, the large-scale radio filaments are localised dynamical structures with $B \sim 1$ mG inside, but no such large field outside. The field in the filaments is regarded as a local and dynamical feature, that is perhaps enhanced by compression but is not representative of the wider galactic centre field.

Faraday rotation measurements generally prefer low values for the central magnetic field,  $B\sim 10\, \mu{\rm G}$ \cite{Roy}.
However, Faraday rotation only probes the component of the magnetic field along the line of sight, while ALP to photon conversion relies on
the field perpendicular to the line of sight.
For a field with random orientation, these will be comparable, but given that there are significant reasons to think that the galactic centre field is strongly poloidal, Faraday rotation estimates of the
parallel magnetic field
cannot be said to give a reliable measurement of the strength of the transverse field.

Magnetic field estimates have also be reported for smaller sub-regions within the central $\sim$ 150 parsec.
Dense molecular clouds are widely argued to support horizontal
 magnetic fields
 of the order of $B\sim $mG \cite{Novak-etal2003,Chuss-etal2003}, with such fields being produced by shearing of the poloidal field by cloud motions or tidal forces.
At a distance of
 0.1 pc from the central black hole,
reference \cite{Eatough2013} found a magnetic field $B > 8\, $mG. However, as the physics
extremely close
 to the supermassive black hole can be expected to differ substantially from that of more distant regions,
 it is unclear what this implies for the magnetic field at distances of 10--100 pc from Sgr A*.

It is clear from the above differences that no definitive statements can be made about the galactic centre magnetic field, and it  is far beyond the scope of this paper to reconcile these different estimates and measurements. Further studies of Faraday rotation measurements and radio continum, as well as improvements in the far infrared/submillimiter polarimetry and Zeeman splitting will help to get a clearer picture for the magnetic field in the galactic centre region.

In this paper, we will primarily focus on the first, maximal, scenario involving a poloidal magnetic field with milligauss field strength.\footnote{We note that magnetic flux conservation implies that the strength of such a field cannot fall off rapidly on a scale of 20--40$\, $pc  above the galactic plane.}
As the $a \to \gamma$ conversion probability scales with $B^2$, this maximises the resulting signal.
We will see that in this maximal scenario, an observable signal from
${\rm DM} \to a \to \gamma$ is just achievable. Due to the $B^2$ dependence of the signal, it follows that we do not need to consider the other (medium and low)
scenarios in detail: they are incapable of generating an observable 3.5 keV line from the galactic centre in the $\rm{DM} \to a \to \gamma$ scenario.

\section{Dark matter  $\to a \to \gamma$ at the centre of the Milky Way}
\label{sec:results}
We are now ready to discuss the characteristics of ALP to photon conversion in the centre of the Milky Way. We use the maximal model for the magnetic field discussed in Section \ref{sec:magneticfield}, i.e.,
\be
\vec{B} = 1\, {\rm mG} \, \hat  z \, ,  \label{eq:Bmodel}
\ee
in the central for $r\lesssim150\, {\rm pc}$ in the galactic plane, with the electron density as given by the galactic centre and thick disk components of the NE2001 model, c.f.~Section \ref{sec:electrondensity}. In particular, we assume $\vec{B} = 1\, {\rm mG} \, \hat  z$ for $\vert x \vert < 150 \, \rm{pc}$ along sight lines within the XMM Newton and Chandra field of views.

 The predicted photon flux per unit steradian is then given by~\eqref{FluxGCgen},
\begin{align}
 \begin{aligned}
  {\cal F}
  &\simeq \frac{1}{4 \pi \tau_{\rm DM}} {\langle P_{a \rightarrow \gamma} \rangle}_{\text{FOV}} \int_{l_\text{GC}}^{\infty} \frac{\rho_{\rm DM} }{m_{\rm DM}} dl \,  . \label{eq:flux}
 \end{aligned}
\end{align}
As sufficiently large conversion probabilities are only  obtained
in the vicinity of the galactic centre,  dark matter decaying between Earth and the galactic centre does not contribute significantly to the observed photon flux. Consequently,  the integration in~\eqref{CDused} is from the galactic centre at $l_\text{GC} = 8.5\, {\rm  kpc}$ and outwards.

The photon flux is sensitive to the dark matter column density. We use an NFW profile \cite{NFW}, given by
\begin{equation}
 \rho_{NFW}(r) = \frac{\rho_{s} r_s}{r \left( 1+ r/r_s \right)^2}\,,
\end{equation}
where $r$ is the distance to the galactic centre, $\rho_{s} = 20.4 \ti 10^6 M_\odot /$ kpc$^3$ (local dark matter density) and $r_s= 10.8$ kpc~\cite{2010A&A...509A..25W}. Averaging over the full \XMM field of view, the column density is given by,
\be
\int_{l_\text{GC}}^{\infty}  \frac{\rho_\text{DM}}{m_\text{DM}} dl \simeq 2.2 \ti 10^{28}~{\rm cm^{-2}} \, ,\label{CDused}
\ee
where we have used $m_{\rm DM} = 7.1$ keV.

Other models of the dark matter density can result in quite different column densities: for the 10 profiles  presented in the Appendix of~\cite{BoyarskyII} (7 NFW, Einasto, ISO and BURK), the column densities
are in the range
\be
\int_{l_\text{GC}}^{\infty}  \frac{\rho_\text{DM}}{m_\text{DM}} dl \simeq 3.5 \ti 10^{27} - 2.9 \ti 10^{28}~{\rm cm^{-2}} \,.\label{CDs}
\ee
This uncertainty should be kept in mind when flux values~\eqref{eq:flux} are calculated in the following.

\subsection{ALP to photon conversion probability}
We now solve equation (\ref{eq:EqofMotion}) to
 derive the axion-photon conversion probability for  ALPs
 with negligible mass, $\Delta_a \ll \Delta_{\gamma}$ and propagating through the galactic centre region of the Milky Way. We will do this in two ways: first, we will solve equation \eqref{eq:soln} analytically by noting that a small-mixing perturbative approximation
is appropriate, and second, we will solve equation \eqref{eq:EqofMotion}
numerically by discretising the evolution of an initially pure ALP-state.

\begin{figure}
\begin{center}
\includegraphics[width=0.96 \textwidth]{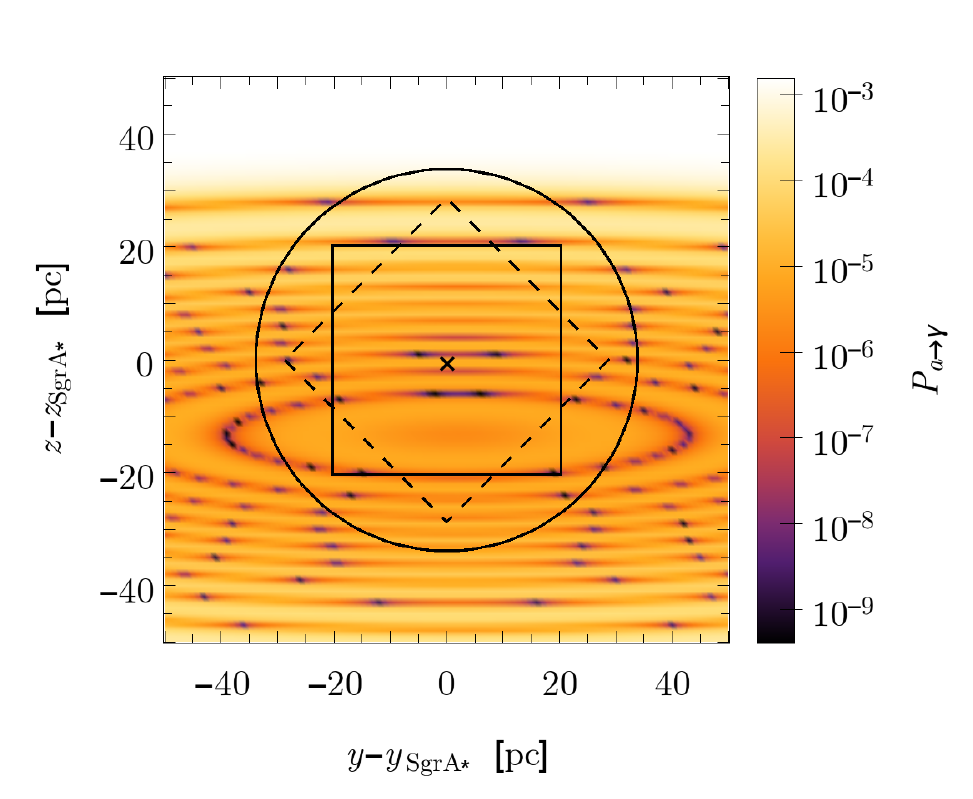}
\end{center}
\caption{The values of $P_{a\to \gamma}$ as a function of the galactocentric coordinates $(y,z)$ according to a numerical simulation of \eqref{eq:EqofMotion} with $M=10^{13}$ GeV.
The outer solid circle indicates the field of view of \XMMs.
The field of view of Chandra is indicated by the solid square (parallel orientation to $y$-axis) and the dashed square ($45^\circ$ orientation to $y$-axis).
As observations are centred on Sgr A* they are slightly offset from $(y,z) =(0,0)$.}
\label{fig:Prob}
\end{figure}

The strength of the ALP-photon mixing can be estimated by considering the ratio,
\bea
\frac{2 B_{\perp} \omega}{M m^2_{eff}} \approx
10^{-3} \ti
\left(\frac{B_{\perp}}{1~{\rm mG}} \right)
\left(\frac{10^{13}~{\rm GeV}}{M} \right)
\left(\frac{3.5~{\rm keV}}{\omega} \right)
  \, , \nonumber
\eea
where we have specialised to $\Delta_a = 0$ and taken $n_e = 10$~cm$^{-3}$.  This suggests that a perturbative, small mixing approximation of the interaction should provide a good approximation to the full solution. To linear order, we find that an initially pure ALP state travelling from $x=-L/2$ to $x=L/2$ may convert into a photon with the probability
\be
P_{a\rightarrow \gamma}(L)=\sum_{i=z,y}\left|\int^{L/2}_{-L/2} dx \, e^{i \varphi(x )}  \Delta_{\gamma_i a}(x)\right|^2 \, ,
\ee
where,
\be
\varphi(x)  = \int^{x}_{-L/2} dx' \Delta_{\gamma}(x') = - \frac{1}{\omega} \int^{x}_{-L/2} dx' \omega_{pl}^2(x') \, .
\ee
For a constant magnetic field, this expression may be further simplified to,
\be
 P_{a\rightarrow \gamma}(L)=  \frac{B_{\perp}^2}{4 M^2} \int_{-L/2}^{L/2} dx_1 \int_{-L/2}^{L/2} dx_2 \cos\left(\frac{2\pi \alpha}{\omega m_e} \int_{x_2}^{x_1} dx' n_e(x') \right) \label{eq:ProbAn1}
 \, .
 \ee
For an electron density with a Gaussian fall-off -- such as  the galactic centre  component of the NE2001 model -- we may perform the $x'$-integral to obtain an argument of the cosine which is proportional to the difference between two error functions with arguments $x_i/L_\text{GC}$ for $i=1,2$, respectively, where $L_\text{GC}$ is defined as in equation \eqref{ne2001gc}. For the maximal magnetic field model of equation \eqref{eq:Bmodel}, we take the magnetic field to be approximately constant  over the central region of the Milky Way, and for the purpose of an analytical estimate, we
 expect that the leading order Taylor expansion of the error functions should well approximate the function over the relevant interval. Explicitly, we approximate,
\be
{\rm Erf}\left(\frac{x_i }{L_\text{GC}}\right) \approx \frac{2}{\sqrt{\pi}} \frac{x_i }{L_\text{GC}} \, ,
\ee
where the subleading corrections appear at cubic order of the argument. With this approximation, we may perform the integrals of equation \eqref{eq:ProbAn1} explicitly to obtain the conversion probability,
\be
 P_{a\rightarrow \gamma}(L)=
\frac{B_{\perp}^2 \omega^2 m_e^2}{ 4\pi^2 \alpha^2 M^2 (n_e^{(0)})^2}
 e^{2 \left(\frac{(y-y_\text{GC})^2 }{L_\text{GC}^2} + \frac{(z-z_\text{GC})^2}{H_\text{GC}^2} \right)}   \sin^2\left( \frac{ \pi \alpha  n_{e}^{(0)} L }{\omega m_e} e^{- \left(\frac{(y-y_\text{GC})^2 }{L_\text{GC}^2} + \frac{(z-z_\text{GC})^2}{H_\text{GC}^2} \right) } \right)  \, .
\label{eq:ProbAn}
\ee
This conversion probability agrees well with that
obtained from a numerical simulations, which we show
in Figure \ref{fig:Prob} for $L=300$ pc, $M=10^{13}$ GeV.

Figure~\ref{fig:Prob} shows a marked suppression of the conversion probabilities at low values of $z$. This arises as 
the conversion probability is sensitive to the difference between the ALP mass and the plasma frequency - and the latter 
is set by the free electron density. High electron densities lead to a large ALP-photon 
mass difference and a suppression of the ALP-photon conversion probabilities.
At larger galactic altitudes, 
the electron density is lowest and the resulting ALP-photon conversion probability is well approximated by the zeroth order expansion of the cosine in equation \eqref{eq:ProbAn1}, giving $ P_{a\rightarrow \gamma}(L) =  B_{\perp}^2 L^2/(4 M^2)$. This explains the apparent constancy of the conversion probability at $z\gtrsim 30\, $pc.

At lower galactic altitude, the
factor
$
n_{e,\text{GC}}(0,y,z) = n_e^{(0)} {\rm exp} \left[- \left(\frac{(y-y_\text{GC})^2 }{L_\text{GC}^2} + \frac{(z-z_\text{GC})^2}{H_\text{GC}^2} \right) \right]
$,
 that encodes the effective line-of-sight electron density for a given path, is too large to justify a zeroth order expansion. On the contrary, the most striking feature of Figure \ref{fig:Prob} are the spatial oscillations in the conversion probability in the $(y,z)$-plane, and these are directly sourced by the varying line-of-sight electron densities, as is clear from the analytical approximation, \eqref{eq:ProbAn}. In addition, the conversion probability scales with an overall factor of $(1/ n_{e,\text{GC}}^{( \rm eff)}(0, y,z))^2$, which explains the suppression of the conversion probability near the galactic plane.

We note that while the galactic centre contribution to the electron density dominates in the very centre of the galaxy, it quickly drops below the contribution from the `thick disc' component of the NE2001 model at large $z$ and $x$. According to this model for the electron density, the thick disc contributes with $n^{\rm thick~disc}_e = 0.1$ cm$^{-3}$ in the central region, and our full model of the electron density should then be $n_e = n_e^{\rm GC} + n_e^{\rm thick~disc}$. In deriving equation \eqref{eq:ProbAn}, we only included the contribution from $n_e^{\rm GC}$ and not that from $n_e^{\rm thick~disc}$. This neglect however, is harmless: for small enough values of $n_e$, the argument of the cosine of equation \eqref{eq:ProbAn1} can be Taylor expanded with the leading order contribution giving $P_{a\rightarrow \gamma} = B_{\perp}^2 L^2/4M^2$. For $n_e \leq 0.1$ cm$^{-3}$, this `small angle' approximation is in good agreement with the numerical solution of equation \eqref{eq:EqofMotion}. We used a full
discretized
simulation of equation \eqref{eq:EqofMotion} with the full NE2001 electron density $n_e = n_e^{\rm GC} + n_e^{\rm thick~disc}$ to obtain the results in Sections \ref{sec:comparison}, \ref{sec:sensitivity} and \ref{sec:external}.

\subsection{Predictions  for XMM-Newton and \ch }
\label{sec:comparison}
The predictions from the dark matter \scen~scenario are now easily obtained by combining \eqref{eq:flux} and the simulation results of \eqref{eq:EqofMotion}.
We first compare the ALP-photon conversion probabilities in the total field of view of \XMM and Chandra. \XMM has a radial total field of view with a radius of $15'$ (although only the inner $14'$ were used in \cite{BoyarskyII}), while Chandra has a square total field of view of $16.8'\ti 16.8'$. Note that the searches for a 3.5 keV line in the galactic centre actually use a smaller field of view than the total one for both \XMM \cite{BoyarskyII} and Chandra~\cite{RiemerSorensen}. As discussed in Section~\ref{sec:searches}, we use 530 arcmin$^2$ for the actual \XMM field of view for the observations in \cite{BoyarskyII, JP}and 240 arcmin$^2$ for the actual Chandra field of view for the observations~\cite{RiemerSorensen}.

The roll-angle, $\alpha_r$, of Chandra was not fixed during the observations considered in \cite{RiemerSorensen}, and hence the exact orientation of the detectors during each observation was not fixed and may well have varied.
As the average conversion probability over the \ch field of view is sensitive to the orientation,
 we here consider two extreme cases as indicated in Figure~\ref{fig:Prob}.
 If the symmetry axes of the \ch field of view are aligned parallel to the $y$ and $z$ coordinate axes, hence $\alpha_r =0^\circ$ in our notation, most of the region with high conversion probability falls outside the field of view.
 A slightly larger average conversion probability can be expected for the
  tilted field of view with $\alpha_r =45^\circ$.

The field of view of the \XMM observations of the galactic centre is a factor of $2.2$ times larger than that of the Chandra observations. Furthermore, as the \XMM observations include a substantial coverage of the $z>20$ pc region where the electron density is suppressed with respect to that of the galactic plane, the ALP-photon conversion probability for \XMM is larger than that of \ch when averaged over the field of view.
For a magnetic field of $B_{\perp} = 1\, $mG which is constant for $|x|< 150\, $pc within the field of view,
the ratio of the averaged conversion probabilities is  given by
\be
\frac{\langle P_{a\to \gamma} \rangle_{XMM}}{\langle P_{a \to \gamma}\rangle_{Chandra}} = \begin{cases} \frac{3.0\ti 10^{-5}}{1.4\ti 10^{-5}} = 2.1 &\mbox{for } \alpha_r =  0^\circ \\
\frac{3.0\ti 10^{-5}}{1.5\ti 10^{-5}} = 2.0 &\mbox{for } \alpha_r =  45^\circ \end{cases} \, .\label{Probratios}
\ee
Combining the larger conversion probability of \XMM with its larger field of view, we find that the expected photon flux ratio between \XMM and \ch is given by,
\be
\frac{{\cal F}_{XMM}}{{\cal F}_{Chandra}} = \begin{cases} 4.6 &\mbox{for } \alpha_r =  0^\circ \\
4.4 &\mbox{for } \alpha_r =  45^\circ \end{cases} \, .
\ee
Such a substantial flux ratio is consistent with a detectable signal  in \XMMs, and a non-detection in Chandra.
For the dark matter column density given in~\eqref{CDused}, and for $\tau_{\rm DM} = 8.0\ti10^{22}\, $s, $M=10^{13}\, $GeV, we find an expected photon flux of,
\bea
{\cal F}_{XMM} &=& 2.9 \ti 10^{-5}~~{\rm photons~ s^{-1} cm^{-2}} \, ,\label{XMMflux}\\
{\cal F}_{Chandra} &=& 6.7 \ti 10^{-6}~~{\rm photons~s^{-1} cm^{-2}} \, , \label{Chandraflux}
\eea
 where we have used $\alpha_r = 45^\circ$ to estimate the Chandra flux.
 The value of
 $\tau_{\rm DM} M^2$ here has been set to match the XMM flux observed by~\cite{BoyarskyII}.

 For comparison, in~\cite{14032370},  the parameter values $\tau_{\rm DM} = 5\ti10^{24}\, $s and $M=10^{13}\, $GeV
 were used, motivated by  the observed flux from galaxy clusters~\cite{Bulbul} and an estimated
   average ALP to photon conversion probability of $\sim 10^{-3}$ for $M = 10^{13} {\rm GeV}$ in the stacked cluster sample.
This value of $10^{-3}$ comes from numerical simulations of the centre of the Coma cluster in \cite{13123947}.
There are however significant uncertainties on this number of $10^{-3}$. Even within Coma, the magnetic field is uncertain to a factor of two, corresponding to a factor of four
uncertainty in conversion probability. It is also probable that conversion probabilities in the centre of the bright cluster Coma are biased high compared to those for
a stacked average of many clusters. We shall also see in Section~\ref{sec:sensitivity} that this ratio of $\tau_{\rm DM, clusters}/\tau_\text{DM, GC} \sim 60$ is highly sensitive to
the assumed electron density profile in the galactic centre, and can vary by a factor of 10 for small changes in the electron scale height.
Therefore, despite the large apparent difference in $\tau_{\rm DM} M^2$,
the observations of a 3.5$\, $keV line from clusters and from the galactic centre may both be explainable as originating from dark matter in the \scen~scenario.\footnote{Decreasing
 $\tau_{\rm DM}$ by an order of magnitude from $5 \ti 10^{24}$s would make the prediction in \cite{14047741} of no observable signal in the Milky Way slightly less strong, but only by
 changing the predicted signal from the general Milky Way halo to two (instead of three) orders of magnitude weaker than that from galaxy clusters.}

Finally, let us apply a masking that restricts the field of view of \XMM to the $\sim 10^{-4}$ conversion probability region $z> 20$ pc, see Figure~\ref{fig:Prob}. The field of view shrinks from $530$ arcmin$^2$ to $90$ arcmin$^2$, but since the field of view averaged conversion probability is significantly larger than for the total field of view of \XMMs, the flux is rather similar to ~\eqref{XMMflux}:
\begin{equation}
 {\cal F}_{XMM}^{z>20 \text{pc}} = 2.1 \ti 10^{-5}~~{\rm photons~ s^{-1} cm^{-2}} \,.
\end{equation}
We see that the \scen~model can reconcile the conflicting results from Chandra and XMM-Newton if the magnetic field in the galactic centre is large enough. In addition, we predict that the clear majority of the XMM-Newton signal will remain when all but the $z > 20$ pc region is masked out, despite the $\sim 80 \%$ reduction in the field of view. This prediction is easily testable and, if confirmed, would be difficult to explain within any other dark matter model.

\subsection{Sensitivity to model parameters}
\label{sec:sensitivity}
In Section \ref{sec:comparison} we noted that for the default model of the electron density, \XMM obtains a higher averaged conversion probability by observing regions at large $z$ where the electron density is suppressed. The significance of this effect is highly dependent on the off-set of the electron density from the galactic centre and the vertical scale height, the values of which appear in \cite{NE2001}
without error bars. Here, we consider the effects of deviations of the vertical offset of the electron density by 100\%, and deviations of the vertical scale height by 50\%
from their default values.

\begin{figure}
\begin{center}
\includegraphics[width=0.5\textwidth]{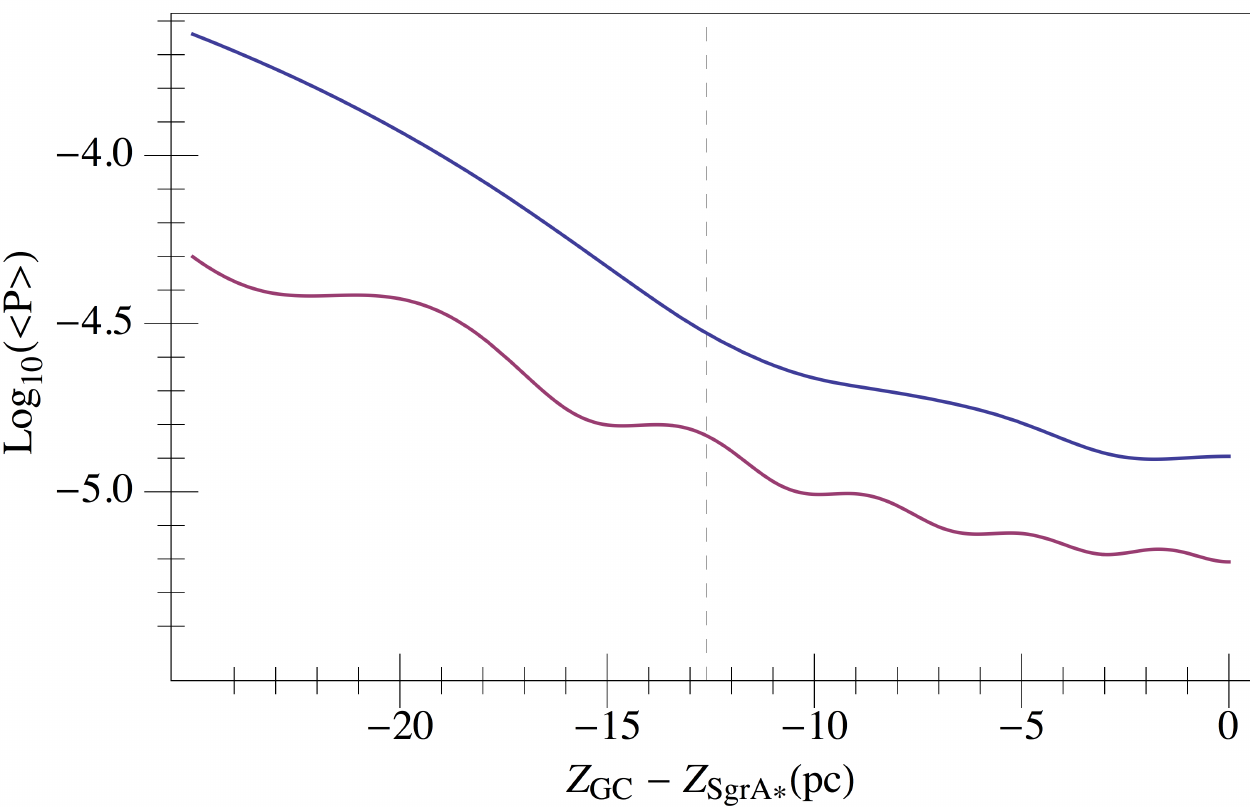}~~~\includegraphics[width=0.5\textwidth]{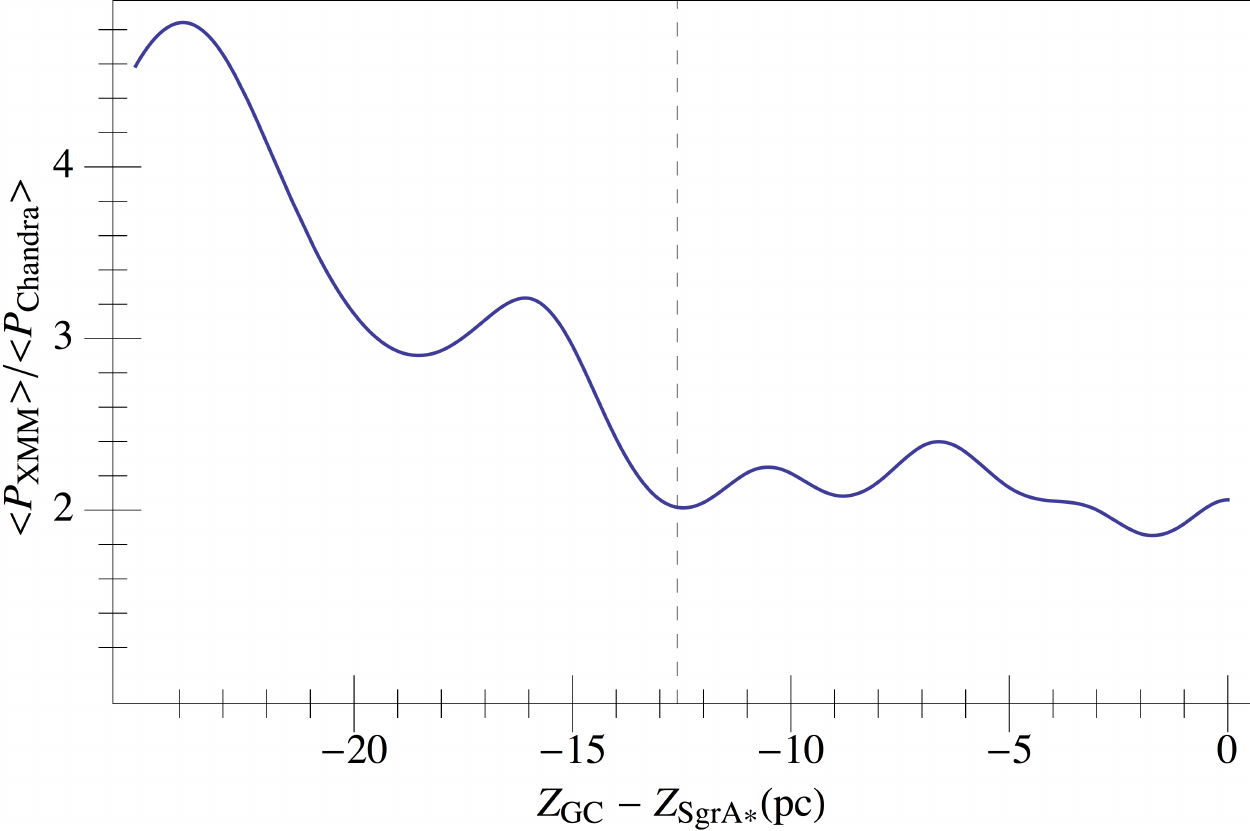}
\end{center}
\caption{Left: The values of ${\rm log}_{10}(\langle P_{a\to \gamma}\rangle)$ for \XMM (blue) and \ch (red) as a function of the off-set of the electron density in the $z$-direction.  Here $M=10^{13}$ GeV. Right: The ratio $\langle P_{a\to\gamma} \rangle_{XMM}/\langle P_{a\to\gamma} \rangle_{Chandra}$ as a function of the off-set. The vertical dashed line indicates the default NE2001 value of $z_\text{GC} - z_{\rm Sgr A^*}=  -12.6$ pc.
}
\label{fig:offset}
\end{figure}
The dependence of the averaged conversion probabilities on the off-set of the electron density are shown in Figure \ref{fig:offset}. Regardless of the off-set, \XMM captures regions at high $z$  with smaller electron densities and thus higher conversion probabilities. The ratio of conversion probabilities -- after averaging over the corresponding field of views for \XMM and \ch -- is ${\cal O}(2$--$5)$. This corresponds to a line flux ratio of ${\cal F}_{XMM}/{\cal F}_{Chandra} \sim {\cal O}(4$--$11)$.

\begin{figure}
\begin{center}
\includegraphics[width=0.45\textwidth]{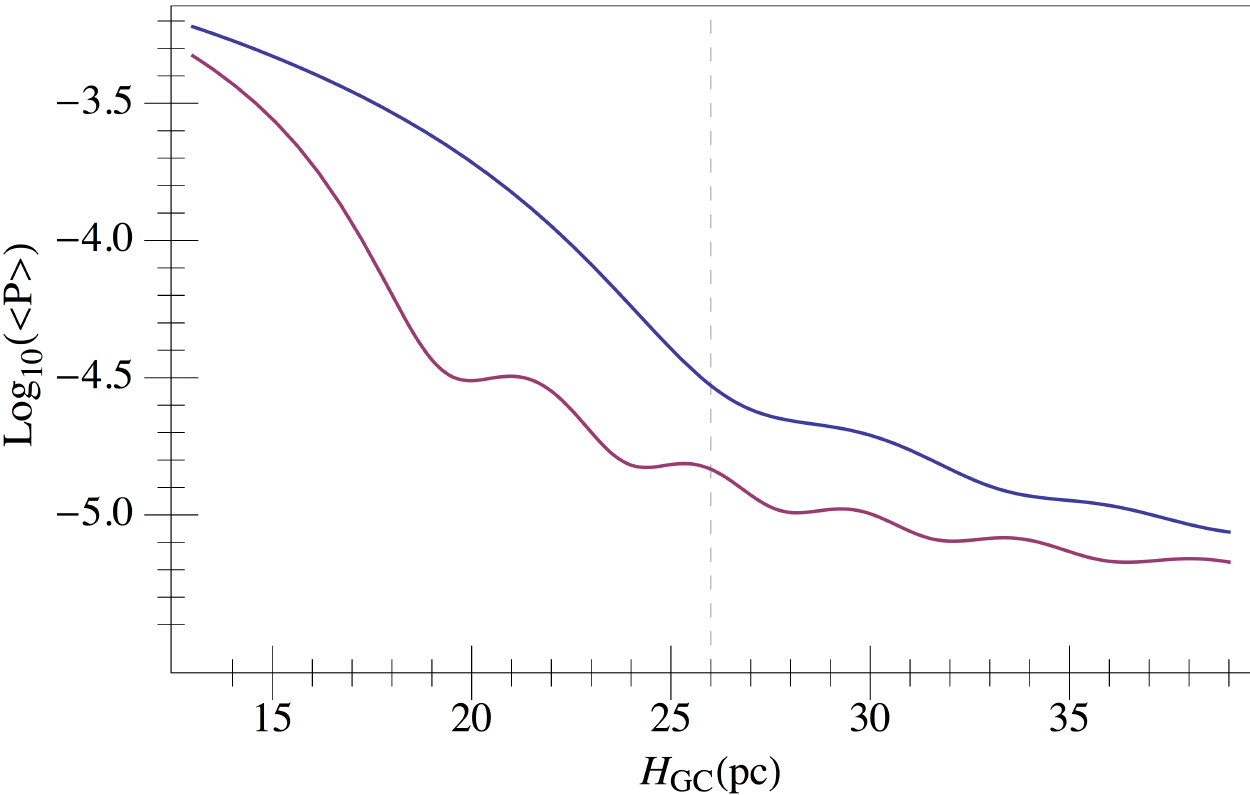}~~~\includegraphics[width=0.45\textwidth]{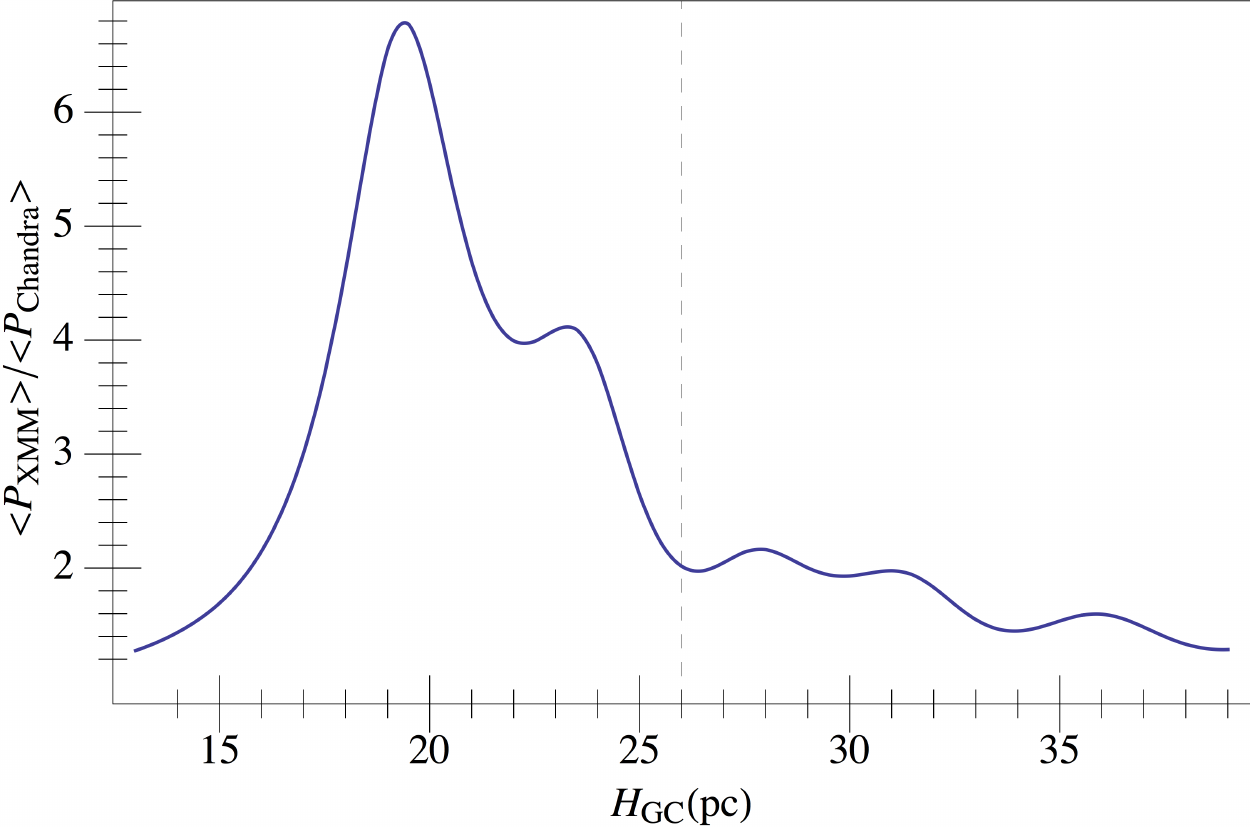}
\end{center}
\caption{Left: The values of ${\rm log}_{10}(\langle P_{a\to \gamma}\rangle)$ for \XMM (blue) and \ch (red) as a function of
the scale height $H_\text{GC}$ of the electron density.  Here $M=10^{13}$ GeV. Right: The ratio $\langle P_{a\to\gamma} \rangle_{XMM}/\langle P_{a\to\gamma} \rangle_{Chandra}$ as a function of the scale height. The vertical dashed line indicates the default NE2001 value of $H_\text{GC} =  26$ pc.}
\label{fig:scaleheight}
\end{figure}

For a scale height $H_\text{GC}$ smaller than the default value of $26$ pc, more of the field of views of both \XMM and \ch capture low electron density regions. This leads to higher averaged conversion probabilities, as is evident from Figure \ref{fig:scaleheight}. The ratio of conversion probabilities are again in the range ${\cal O}(1$--$6)$. We note that the predicted conversion probability is substantially increased by small increases in the electron density offset or small decreases in the scale height. This would correspond to a much lower predicted value of $\tau_{\rm DM} M^{2}$. While the off-set in the $y$-direction and the in-plane suppression length $L_\text{GC}$ are also factors of the electron density, a  variation of their values by an ${\cal O}(1)$ factor has a negligible effect on the conversion probabilities.

As mentioned above, only the uppermost end of the observational estimates for the galactic centre field lead to an obervable signal. To quantify this, we also consider the following magnetic field model: An ambient magnetic field of 0.1 m$G$ with $N_{rf}$ cylinders with radius 0.5 pc stretching from $z=-100$ pc to $z=100$ pc with magnetic field 1 m$G$ corresponding to observed radio filaments. We organize the cylinders on a grid in the $x$-$y$ plane with gridlength 3 pc over a disk with radius 50 pc. The averaged conversion probabilities for this magnetic field model are
\be
\frac{\langle P_{a\to \gamma} \rangle_{XMM}}{\langle P_{a \to \gamma}\rangle_{Chandra}} = \begin{cases} \frac{3.0 \ti 10^{-7}}{1.4 \ti 10^{-7}}=2.1 &\mbox{for } \alpha_r =  0^\circ \\
\frac{3.0 \ti 10^{-7}}{1.5 \ti 10^{-7}}= 2.0&\mbox{for } \alpha_r =  45^\circ \end{cases}\,,
\ee
while the predicted fluxes become
\begin{align}
 \begin{aligned}
  {\cal F}_{XMM} &= 2.9 \ti 10^{-7}~~{\rm photons~ s^{-1} cm^{-2}} \, ,\\
{\cal F}_{Chandra} &= 6.7 \ti 10^{-8}~~{\rm photons~s^{-1} cm^{-2}} \, ,
 \end{aligned}
\end{align}
where again we have used $\alpha_r = 45^\circ$ to calculate the Chandra flux. As expected, these fluxes are a factor of 100 lower than in the case of a pervasive 1 mG field, corresponding to reducing the magnetic field strength by a factor of 10. The radio filaments are much too narrow to contribute to the conversion probability. In this field model, the signal from the Milky Way centre cannot be explained by \scen.

\section{Searching for the 3.5 keV line in other galaxies} \label{sec:external}
In this section, we discuss the search for the 3.5 keV X-ray line in galaxies other than the Milky Way and the inferred constraints on dark matter models.
We argue that currently published studies of the X-ray line in other galaxies are consistent with the \scen~scenario, and we also provide a list of target galaxies
in the \XMM and \ch archives
with significant exposures for which a detection of the 3.5 keV signal is more likely in this scenario.

\subsection{Observational hints and constraints}
The first search for the 3.5 keV line in a galaxy was that of \cite{Boyarsky}, who detected the line in the combined \XMMs~spectrum of the central region of M31.  We reviewed this result in Section \ref{sec:intro}.

In \cite{Anderson}, stacked \ch and \XMM observations of  galaxies were used to constrain the proposed sterile neutrino origin of the line. For \ch and \XMM respectively, archived data from a sample of 81 and 89 galaxies with a total exposure of 15.0 Ms and 14.6 Ms was considered.
The stacking was made so as to optimise sensitivity to lines from decaying dark matter by minimising the X-ray background.
To avoid an ICM background, no galaxies in clusters or groups with temperature $T\gtrsim 1$ keV were included in the sample.
 In addition, to avoid emission from the inner-most regions of galaxies, data was only extracted from an annular region within the radius $r \in [0.01 R_{\rm vir.}, R_{\rm vir}]$, where  $R_{\rm vir.}$ denotes  the estimated virial radius of each galaxy.
 The resulting X-ray spectra were then argued to be dominated by instrumental background. To avoid prominent instrumental lines, the search was then restricted to the ranges $2.6$--$5.2$ keV for \ch and $2.4$--$6.2$ keV for \XMMs.
This background was fitted by slowly varying smoothing splines, above which any potential  additional emission line would appear as a  localised residual of the fit.  To constrain the presence of a line at 3.57 keV, a zero-width ``Gaussian'' component was added to the background spline, and the resulting fit was compared to that without the added line. When the amplitude of the added line was fixed to the central value inferred from \cite{Bulbul}, the line was found to be ruled out at 4.4$\sigma$ for the \ch spectrum and at 11.8$\sigma$ for the \XMM spectrum. When the amplitude of the line was left to freely vary, the preferred amplitude was found to be consistent with zero,
thus favouring the model without the additional line.

In \cite{Malyshev}, a search for the 3.5 keV line in stacked \XMM data from $0.6$ Ms of observations of 8 dwarf spheroidal galaxies was preformed.  As the interstellar medium of dwarf spheroidals does not emit X-rays, the keV-range background is cleaner than that of clusters. The stacked spectra were fitted with models for the astrophysical and instrumental backgrounds, and a narrow line at 3.55 keV was added to these models and shown not to improve the fit. To constrain the dark matter origin of the line, the contribution of a possible 3.55 keV line from both the Milky Way halo and the dwarf spheroidals was considered, and for sterile neutrinos, a mixing angle of the magnitude inferred in \cite{Bulbul} was shown to be excluded at 3.3$\sigma$ or 4.6$\sigma$ depending on assumptions.

These results of \cite{Anderson, Malyshev} provide strong constraints on scenarios with dark matter decaying directly to photons.
We now explain how in the scenario of \scen, no detectable signal would arise for these searches.

\subsection{Predictions for the  \scen~model  in other galaxies}

In contrast to scenarios in which dark matter decays or annihilates into photons, the observed strength of the X-ray line in the \scen~scenario depends on the magnetic field along the line of sight to the point of decay. The prospects of observing a signal from galaxies therefore depend on the structure and strength of the magnetic field in galaxies, which differs strongly between different morphological types of galaxies (for a review, see e.g.,~\cite{Beck:2012}).

While the origin  of galactic magnetic fields is not well known, enhancement of small seed magnetic fields through dynamo mechanisms provide a plausible explanation for their development. The magnetic fields in galaxies can be split into contributions from three components. The random field is a short scale, tangled magnetic field, with coherence length $\sim100\, {\rm pc}$ (the typcial size of a supernova outflow). The random field may be enhanced to $\mu$G strengths by turbulence within the galaxy. Many spiral galaxies also have a regular field that is coherent over large distances. These may be generated by a mechanism like the mean-field dynamo, which produced spiral magnetic fields coherent over significant distances through differential rotation. Indeed, the magnetic field in the disc of a spiral galaxy generally follows the pattern of the spiral arms. Finally, galaxies may have striated fields, in which the direction of the field is coherent over large distances, but the sign of the field is randomised
with a short coherence length. Striated fields may be generated by the levitation of hot plasma bubbles and their associated random fields, or may arise from the random field by differential rotation. In most cases, it is only the regular field component that leads to significant $a \to \gamma$ conversion -- as the regular fields have the largest coherence
scales and $P(a \to \gamma) \propto L^2$.

The observed strong correlation between the total radio continuum emission at centimetre wavelengths and the far-infrared luminosity of star forming galaxies, i.e.,~the `radio-infrared correlation', can be interpreted as a correlation between the  field strength of the turbulent component and the star formation rate of the galaxy. Regions with high star formation tend to have strong turbulent magnetic fields (indeed, the highest turbulent magnetic fields are observed in highly star-forming starburst galaxies). The \emph{regular} component however is not believed to be positively correlated with the star formation rate: in spiral galaxies the ordered magnetic field, combining the regular and striated fields, is in fact strongest in the inter-arm regions.

Dwarf spheroidal galaxies lack both ordered rotation and significant star formation, and consistent with expectations from the known dynamo mechanism, do not support significant magnetic fields \cite{Beck:2012}. Taking a turbulent magnetic field $B \sim 1 \, \mu {\rm G}$ (as in \cite{1301.5306}) with coherence length $L \sim 100 \, {\rm pc}$ for a dSph of diameter 1 kpc, the small angle approximation gives,
\be
P_{a \to \gamma, dSph} \sim 2.3 \ti 10^{-9} \left( \frac{10^{13} {\rm GeV}}{M} \right)^2 \, .
\ee
In the \scen~scenario, decaying dark matter in dwarf spheroidal galaxies will give rise to a 3.5 keV ALP line, but no associated photon line will be observable.
The dominant contribution to the flux from dSphs is therefore from conversion in the Milky Way, which as shown in \cite{14047741}
is too low for an observable signal.

Spiral galaxies tend to support ordered (regular and striated) fields. The ordered fields are strongest between spiral arms where typical values are $B_{\rm ordered} \sim $10--15 $\mu$G
\cite{Beck:2012}. Note that this value includes the contribution from both the regular and the striated fields, while we are typically only concerned with the regular field.
 Within the spiral arms, the magnetic field
 is mostly tangled and randomly oriented.
M31 is unusual in that it has an unusually coherent regular magnetic field $B_{\rm reg} \sim 6\, \mu$G whose strength remains constant across the spiral arms. Spiral galaxies, like the Milky Way, may also support magnetic fields in the halo surrounding the disc. Star burst galaxies can support very strong magnetic fields, however, these tend to be tangled over short scales.

In the \scen~scenario, we expect no line to be observable from elliptic and irregular galaxies which lack large-scale regular magnetic fields \cite{14032370}.
Spiral magnetic fields may in principle give rise to an observable signal if the regular magnetic field is sufficiently strong along the path of an ALP arising from dark matter decay. As ALP-photon conversion is suppressed by the plasma frequency, a stronger signal is expected from regions with small electron density and significant regular magnetic field.
This suggests the inter-arm regions of typical spiral galaxies will typically contribute more to the photon line than the arm regions. Moreover, edge-on galaxies for which a large fraction of the ALPs travel through a significant fraction of the disc magnetic field should yield a larger signal than face-on spiral galaxies. The magnetic field direction follows the direction of the spiral arms. Therefore, for paths within a few kpc of the centre the field is generally transverse to the line of sight, whereas for paths further from the centre the field becomes parallel to the line of sight and so does not contribute to conversion. We therefore predict that the line
flux is much stronger for on-centre observations, but may indeed be unobservable off-centre.

To quantify this, we simulated the expected signal for
hypothetical galaxies with electron density and magnetic field such as those of the Milky Way and M31, observed at inclination angle $\theta_i$. We will refer to these models `Milky Way-like' and `M31-like'.
We assumed that these galaxies were located a
 1 Mpc from Earth and observed with
  a circular field of view with a $15'$ radius pointed at the centre of a galaxy
 (so that the central 4.4 kpc of the galaxy are included in the observation).

 For the Milky Way-like galaxy we used the recent magnetic field model by \cite{Farrar1} with the central 1 kpc sphere (not considered in \cite{Farrar1}) filled in with a $5 \, \mu\rm{G}$ poloidal field with an exponential vertical scale height of 1 kpc. We use the electron density given in the thick and thin disc components of \cite{NE2001} (imposing a minumum value of $n_{e} = 10^{-7} \, \rm{cm}^{2}$) and a NFW dark matter distribution \cite{NFW} with the parameters given in \cite{1305.5597}. We note that the Milky Way magnetic field includes a significant halo component in addition to the disk component, whereas there is no evidence for such a halo component in M31.

 For the M31-like galaxy, based on \cite{FletcherBeck} we assume a constant azimuthal field of $5 \, \mu {\rm G}$ in the disk cut off at a cylindrical radius of 20 kpc. We assume an exponential fall off above and below the disk with a scale height of 2 kpc. This is clearly a vastly simplifed representation of the true field in M31,
underestimating the field in the centre and overestimating the field on the outskirts, but is sufficient to predict the qualitative relationship between inclination angle and flux. We use the electron density
\be
n_{e} = 0.09 \, \rm{cm}^{-3} \times e^{-\frac{\left( r - 3.7 \, \rm{kpc} \right) ^{2}}{160 \, \rm{kpc}^{2}}} \times \rm{sech}^{2} \left( \frac{z}{0.14 \, \rm{kpc}} \right)\,.
\ee
This is an adapted version of the thin disk component of \cite{NE2001}, chosen by considering the electron density values given in \cite{FletcherBeck}. For the Milky Way-like case, we impose a minimum electron density of $n_{e} = 10^{-7} \, \rm{cm}^{2}$. We assume an NFW dark matter distribution with parameters from \cite{astroph0501240}.

\begin{figure}
\begin{center}
\includegraphics{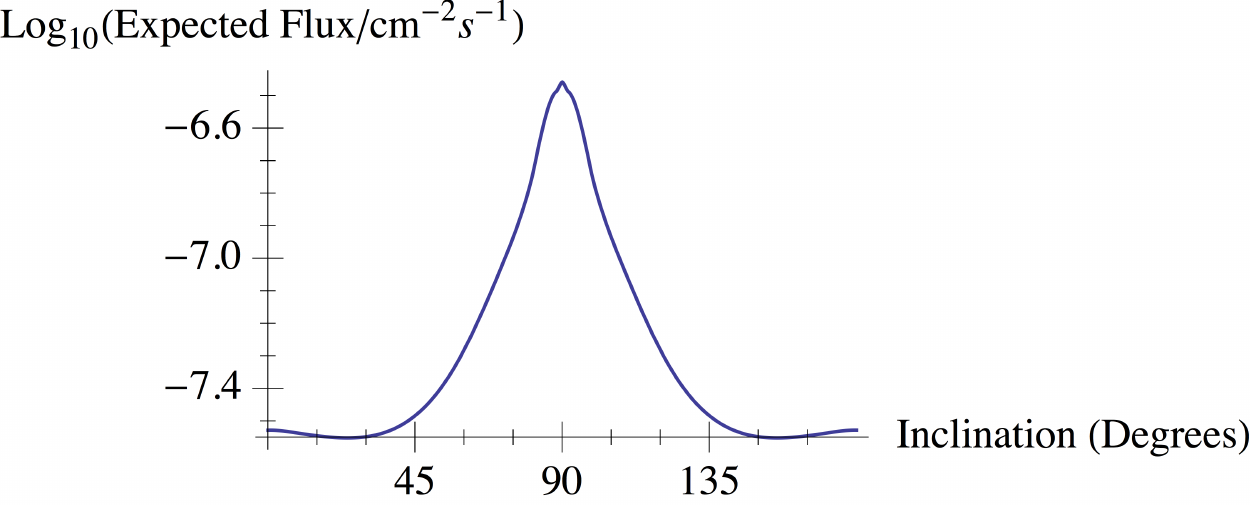}
\end{center}
\caption{Expected flux vs inclination angle for an M31-like Galaxy}
\label{fig:M31Like}
\end{figure}

\begin{figure}
\begin{center}
\includegraphics{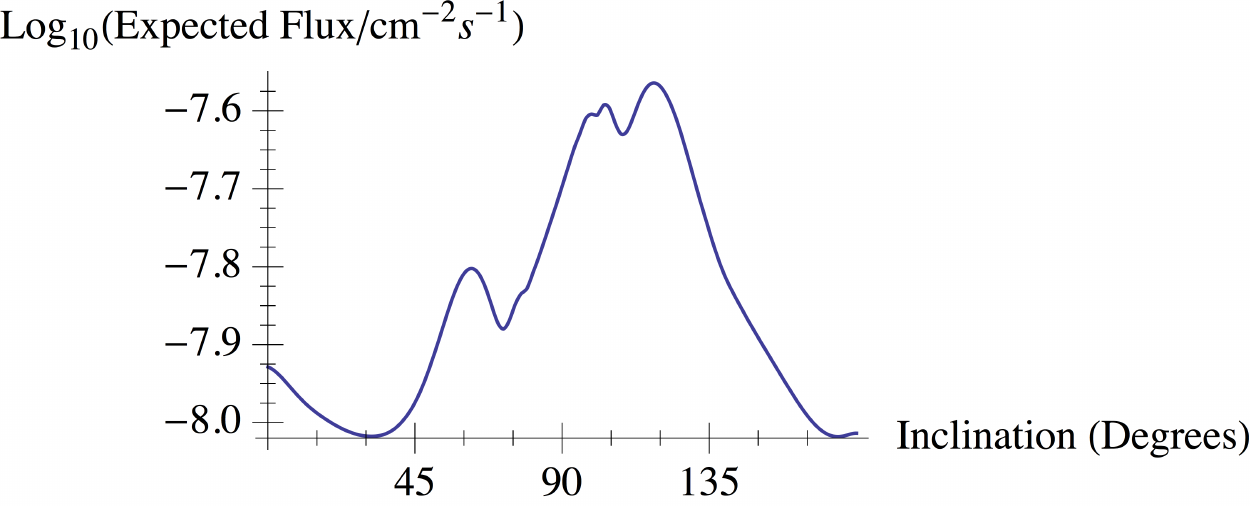}
\end{center}
\caption{Expected flux vs inclination angle for a Milky Way-like galaxy}
\label{fig:MWLike2}
\end{figure}

For the M31-like galaxy in Figure \ref{fig:M31Like}, the peak flux is expected at inclination angle $\theta_i = 90^{\circ}$ (edge-on). The flux for such an edge-on galaxy is over 10 times the flux for an equivalent galaxy with $\theta_i = 0^{\circ}$ (face-on). Note that in this case the magnetic field model used is symmetric above and below the disc, and so the expected flux will be symmetric around $\theta_i = 90^{\circ}$. For the Milky Way-like galaxy in Figure \ref{fig:MWLike2}, the expected flux is lower primarily due to the smaller and less coherent field. Furthermore, rotating the galaxy from $\theta_i = 0^{\circ}$ to $\theta_i = 90^{\circ}$ only increases the flux by a factor of $\sim 3$. This is due to the significant halo component of the Milky Way field. The halo component of the Milky Way field is not symmetric above and below the disc, and so the expected flux is not symmetric about $\theta_i = 90^{\circ}$, and in fact the maximum flux occurs at an inclination angle somewhat above $\theta_i = 90^{
\circ}$.

 In a search for the \scen~model, Figures \ref{fig:M31Like} and \ref{fig:MWLike2} make it clear that we should consider a stacked sample of close to edge-on spiral galaxies. To demonstrate that such a search is feasible, in Appendix \ref{sec:table} we provide a list of spiral galaxies with an apparent diameter of at least $1'$ with $\theta_i \geq 65^{\circ}$ that all have significant exposures in either the \XMM or \ch archives.\footnote{We consider galaxies for which the  sum of \XMM and \ch  exposure is at least 5 ks.}  There are 125 and 143 such galaxies in the \ch and \XMM archives with total raw exposures of 7.1 Ms and 8.7 Ms, respectively. A significant fraction of  these exposure times may well be used to search for the 3.5 keV line.
In a search optimised for the \scen~model, the masking of the field of view would differ from that used in \cite{Anderson}.
For distant galaxies, the whole galaxy might fit in the field of view, whereas we only expect an observable signal from the central region.
In our case, the outer regions of galaxies should be masked and observations instead focused on the central regions.
If it were observationally possible, galaxies with high regular magnetic fields should be preferred. However for more distant galaxies their regular magnetic field
is unknown and this would not be practical.\footnote{We also note that the lack of precise knowledge on the galactic magnetic fields imply that one cannot decisively
rule out the model based on such a search; a definitive exclusion would require knowledge of the magnetic fields.}

\section{Conclusions}
The \scen~scenario
represents an attractive and testable proposal to explain the 3.5 keV line emission, assuming it is of dark matter origin.
At the current time this scenario is consistent with all observations, and can explain discrepancies that cannot be accounted for in models of dark matter
directly decaying or annihilating into photons.
In this paper we have further elucidated the phenomenology of this scenario.

In the galactic centre region, we have studied the conditions under which this scenario can generate a 3.5 keV line of the strength observed in
\cite{JP, BoyarskyII}.\footnote{As discussed in both \cite{JP} and \cite{BoyarskyII}, it is of course possible that the galactic centre line is simply
an astrophysical K XVIII line and there is no dark matter signal.}
 This turns out to be just possible -- provided the magnetic field in the galactic centre is at the highest end of observational estimates.
 It also requires the `average' ALP-to-photon conversion probability for the 73 cluster sample of \cite{Bulbul} to be slightly smaller than assumed
 in \cite{14032370}. In this case, the scenario
generates a highly distinctive morphology, in which the signal is highly suppressed within 20 pc of the galactic plane. This morphology can be easily tested by re-analysing
the data used in \cite{JP, BoyarskyII} and masking the region close to the galactic plane.

We have also considered samples of distant galaxies, and have further quantified the qualitative statement in \cite{14032370} that edge-on spiral galaxies
are the most attractive galaxies for dark matter searches in this scenario. To this end we have also provided a list of target galaxies with significant archival observational
time in the XMM-Newton and Chandra archives.

\section*{Acknowledgments}

JC thanks the Royal Society for a University Research Fellowship. JC, FD, DM, MR are funded by the ERC Starting Grant `Supersymmetry Breaking in String Theory'. PDA is supported by CONICYT Beca Chile 74130061.
Portions of this work have been presented at the `Particle Cosmology after Planck' workshop at DESY in September 2014.
We thank Stephen Angus, Alexey Boyarsky, Thorsten Bringmann, Jeroen Franse, Carlos Frenk, Tesla Jeltema, Andrew Powell, Stefano Profumo, Signe Riemer-S\o rensen, Oleg Ruchayskiy for discussions and correspondence.
DM is grateful to Birzeit University for kind hospitality while finishing the paper.

\appendix
\section{List of nearly edge-on spiral galaxies with long X-ray exposures}
\label{sec:table}
We here list a set of galaxies with a large apparent diameters,\footnote{Using the function {\it logdc} of the Hyperleda database,  {\tt http://leda.univ-lyon1.fr},  \cite{HyperLeda}. } which have inclination angles $\theta_i \geq 65^{\circ}$ and exposures with \XMM and \ch of at least 5 ks. These galaxies would constitute natural targets for a search for the 3.5 keV line from the \scen~scenario. In compiling this list, we only considered observations centred within $2'$ of the target galaxy. Here, $n_{\rm CXO}$ and $n_{\rm XMM}$ denotes the number of such observations available in the archives of \ch and \XMM, respectively.

\begin{small}
\begin{center}
\begin{longtable}{llccccc}
\caption{List of nearly edge-on spiral galaxies with  long X-ray exposures}\\
\tt{Galaxy} & \tt{Type} & $\theta_i$ & $n_{\rm CXO}$ & $t_{\rm CXO}~{\rm [ks]}$ & $n_{\rm XMM}$ & $t_{\rm XMM}~{\rm [ks]}$ \\
\toprule
\endfirsthead
\multicolumn{7}{c}%
{\tablename\ \thetable\ -- \textit{Continued}} \\
 \tt{Galaxy} & \tt{Type} & $\theta_i$ & $n_{\rm CXO}$ & $t_{\rm CXO}~{\rm [ks]}$ & $n_{\rm XMM}$ & $t_{\rm XMM}~{\rm [ks]}$ \\
\toprule
\endhead
\hline \multicolumn{7}{c}{\textit{Continued on next page}} \\
\endfoot
\hline
\endlastfoot
 \tt{ESO602-031} & \tt{SBb} & 70.8 & 1 & 5.0& 2 & 20.9 \\
 \tt{IC2163} & \tt{Sc} & 78.2 & 2 & 40.2 & 1 & 14.9 \\
 \tt{IC2560} & \tt{SBb} & 65.6 & 2 & 65.6 & 3 & 64.9 \\
 \tt{IC2574} & \tt{SABm} & 83.0 & 1 & 11.4 & 2 & 61.0\\
 \tt{IC2810} & \tt{SBab} & 75.2 & 1 & 15.0& 5 & 331.0\\
 \tt{NGC0224} & \tt{Sb} & 72.2 & 105 & 939.9 & 1 & 96.3 \\
 \tt{NGC0253} & \tt{SABc} & 90.0 & 6 & 159.8 & 1 & 14.8 \\
 \tt{NGC0520} & \tt{Sa} & 75.7 & 1 & 50.0& 8 & 371.0\\
 \tt{NGC0625} & \tt{SBm} & 90.0 & 1 & 61.1 & 3 & 130.0\\
 \tt{NGC0660} & \tt{Sa} & 78.7 & 4 & 58.2 & 2 & 151.6 \\
 \tt{NGC0891} & \tt{Sb} & 90.0 & 3 & 174.0& 43 & 977.1 \\
 \tt{NGC0931} & \tt{Sbc} & 81.3 & 1 & 5.0& 3 & 102.0\\
 \tt{NGC1808} & \tt{Sa} & 83.9 & 1 & 43.4 & 10 & 273.0\\
 \tt{NGC2683} & \tt{Sb} & 82.8 & 1 & 1.8 & 2 & 69.7 \\
 \tt{NGC2798} & \tt{SBa} & 83.4 & 1 & 5.2 & 1 & 29.5 \\
 \tt{NGC2799} & \tt{SBm} & 90.0 & 1 & 5.2 & 1 & 24.6 \\
 \tt{NGC2841} & \tt{Sb} & 65.2 & 2 & 30.4 & 1 & 31.4 \\
 \tt{NGC2903} & \tt{SABb} & 67.1 & 1 & 94.8 & 1 & 34.2 \\
 \tt{NGC2992} & \tt{Sa} & 90.0 & 1 & 50.2 & 1 & 40.7 \\
 \tt{NGC3034} & \tt{Scd} & 76.9 & 23 & 812.5 & 1 & 11.2 \\
 \tt{NGC3079} & \tt{SBcd} & 90.0 & 1 & 26.9 & 1 & 14.9 \\
 \tt{NGC3221} & \tt{Sc} & 65.7 & 1 & 19.6 & 1 & 58.2 \\
 \tt{NGC3396} & \tt{SBm} & 90.0 & 1 & 19.8 & 2 & 47.5 \\
 \tt{NGC3623} & \tt{SABa} & 90.0 & 1 & 1.8 & 2 & 49.7 \\
 \tt{NGC3627} & \tt{SABb} & 67.5 & 2 & 52.0& 2 & 100.8 \\
 \tt{NGC3628} & \tt{Sb} & 79.3 & 2 & 60.5 & 1 & 34.2 \\
 \tt{NGC3877} & \tt{Sc} & 83.2 & 5 & 121.6 & 1 & 47.3 \\
 \tt{NGC3972} & \tt{SABb} & 81.5 & 1 & 10.1 & 2 & 100.0\\
 \tt{NGC4013} & \tt{Sb} & 90.0 & 2 & 85.1 & 1 & 26.6 \\
 \tt{NGC4039} & \tt{SBm} & 71.2 & 7 & 425.2 & 1 & 54.8 \\
 \tt{NGC4224} & \tt{Sa} & 75.8 & 1 & 2.0& 1 & 44.9 \\
 \tt{NGC4244} & \tt{Sc} & 65.4 & 1 & 49.8 & 1 & 39.7 \\
 \tt{NGC4258} & \tt{SABb} & 68.3 & 4 & 46.0& 1 & 18.7 \\
 \tt{NGC4388} & \tt{Sb} & 90.0 & 2 & 48.2 & 1 & 19.5 \\
 \tt{NGC4395} & \tt{Sm} & 90.0 & 4 & 79.3 & 4 & 162.6 \\
 \tt{NGC4490} & \tt{SBcd} & 79.0& 3 & 98.9 & 1 & 14.4 \\
 \tt{NGC4565} & \tt{Sb} & 90.0 & 2 & 62.8 & 1 & 13.1 \\
 \tt{NGC4569} & \tt{SABa} & 70.8 & 2 & 41.4 & 2 & 50.5 \\
 \tt{NGC4631} & \tt{SBcd} & 90.0 & 1 & 60.0& 1 & 14.4 \\
 \tt{NGC4666} & \tt{SABc} & 69.6 & 1 & 5.0& 1 & 17.6 \\
 \tt{NGC4698} & \tt{Sab} & 73.4 & 1 & 30.4 & 3 & 30.8 \\
 \tt{NGC4945} & \tt{SBc} & 90.0 & 3 & 249.9 & 1 & 39.9 \\
 \tt{NGC5005} & \tt{SABb} & 77.0& 1 & 5.0& 3 & 92.1 \\
 \tt{NGC5170} & \tt{Sc} & 90.0 & 1 & 33.4 & 1 & 14.2 \\
 \tt{NGC5253} & \tt{SBm} & 85.3 & 3 & 194.0& 1 & 32.1 \\
 \tt{NGC5506} & \tt{Sa} & 90.0 & 2 & 10.2 & 1 & 23.7 \\
 \tt{NGC5746} & \tt{SABb} & 90.0 & 1 & 37.3 & 1 & 66.0\\
 \tt{NGC5775} & \tt{SBc} & 83.2 & 1 & 59.0& 1 & 71.6 \\
 \tt{NGC5793} & \tt{Sb} & 78.0 & 2 & 40.8 & 1 & 18.0\\
 \tt{NGC5907} & \tt{SABc} & 90.0 & 2 & 30.1 & 1 & 25.6 \\
 \tt{NGC6118} & \tt{Sc} & 68.7 & 1 & 8.1 & 2 & 88.4 \\
 \tt{NGC7090} & \tt{Sc} & 90.0 & 2 & 57.4 & 1 & 25.8 \\
 \tt{NGC7212} & \tt{Sb} & 76.4 & 1 & 20.2 & 2 & 29.7 \\
 \tt{NGC7331} & \tt{Sbc} & 70.0 & 1 & 30.1 & 2 & 75.3 \\
 \tt{NGC7582} & \tt{SBab} & 68.2 & 2 & 19.6 & 2 & 38.0\\
 \tt{NGC7590} & \tt{Sbc} & 69.4 & 1 & 30.1 & 2 & 38.7 \\
 \tt{NGC7771} & \tt{Sa} & 66.7 & 1 & 17.2 & 7 & 246.4 \\
 \tt{PGC014370} & \tt{Sc} & 78.8 & 2 & 8.1 & 1 & 17.9 \\
 \tt{PGC037477} & \tt{Sb} & 76.8 & 4 & 100.7 & 1 & 19.6 \\
 \tt{PGC044990} & \tt{Sc} & 83.8 & 1 & 15.1 & 1 & 48.9 \\
 \tt{PGC046710} & \tt{SBb} & 81.2 & 1 & 7.2 & 2 & 31.5 \\
 \tt{PGC093080} & \tt{Sc} & 70.6 & 1 & 50.0& 1 & 16.1 \\
 \tt{PGC1110773} & \tt{Sab} & 71.8 & 8 & 233.0& 1 & 13.1 \\
 \tt{UGC12915} & \tt{SBc} & 73.4 & 1 & 40.0& 1 & 15.0\\
 \tt{ESO069-006} & \tt{SBb} & 76.4 & 1 & 14.7 & \tt{} & \tt{} \\
 \tt{ESO137-001} & \tt{SBc} & 66.2 & 1 & 141.9 & \tt{} & \tt{} \\
 \tt{ESO244-030} & \tt{SABb} & 68.6 & 1 & 10.1 & \tt{} & \tt{} \\
 \tt{ESO293-034} & \tt{SBc} & 74.6 & 1 & 10.0& \tt{} & \tt{} \\
 \tt{ESO415-029} & \tt{Sbc} & 77.3 & 3 & 79.6 & \tt{} & \tt{} \\
 \tt{ESO430-020} & \tt{SABc} & 70.3 & 1 & 10.1 & \tt{} & \tt{} \\
 \tt{ESO432-006} & \tt{Sbc} & 78.8 & 1 & 16.3 & \tt{} & \tt{} \\
 \tt{IC0564} & \tt{Scd} & 77.2 & 1 & 15.2 & \tt{} & \tt{} \\
 \tt{NGC0024} & \tt{Sc} & 70.1 & 1 & 43.8 & \tt{} & \tt{} \\
 \tt{NGC0055} & \tt{SBm} & 90.0 & 1 & 9.7 & \tt{} & \tt{} \\
 \tt{NGC0988} & \tt{Sc} & 68.7 & 1 & 5.3 & \tt{} & \tt{} \\
 \tt{NGC1589} & \tt{Sab} & 80.3 & 1 & 10.2 & \tt{} & \tt{} \\
 \tt{NGC1741} & \tt{Sm} & 70.7 & 1 & 36.0& \tt{} & \tt{} \\
 \tt{NGC2552} & \tt{SABm} & 68.0& 1 & 8.0& \tt{} & \tt{} \\
 \tt{NGC2748} & \tt{Sbc} & 68.1 & 1 & 30.0& \tt{} & \tt{} \\
 \tt{NGC2770} & \tt{SABc} & 82.3 & 1 & 18.1 & \tt{} & \tt{} \\
 \tt{NGC2783B} & \tt{Sb} & 90.0 & 1 & 18.2 & \tt{} & \tt{} \\
 \tt{NGC3190} & \tt{Sa} & 87.8 & 1 & 20.1 & \tt{} & \tt{} \\
 \tt{NGC3198} & \tt{Sc} & 77.8 & 1 & 62.4 & \tt{} & \tt{} \\
 \tt{NGC3287} & \tt{SBd} & 75.3 & 1 & 19.1 & \tt{} & \tt{} \\
 \tt{NGC3556} & \tt{SBc} & 67.5 & 1 & 60.1 & \tt{} & \tt{} \\
 \tt{NGC3621} & \tt{SBcd} & 67.5 & 1 & 23.4 & \tt{} & \tt{} \\
 \tt{NGC3683} & \tt{SBc} & 68.8 & 3 & 139.0& \tt{} & \tt{} \\
 \tt{NGC3718} & \tt{Sa} & 66.5 & 1 & 5.4 & \tt{} & \tt{} \\
 \tt{NGC4088} & \tt{SABc} & 71.2 & 1 & 20.1 & \tt{} & \tt{} \\
 \tt{NGC4178} & \tt{Scd} & 90.0 & 1 & 40.0& \tt{} & \tt{} \\
 \tt{NGC4216} & \tt{SABb} & 90.0 & 1 & 5.3 & \tt{} & \tt{} \\
 \tt{NGC4217} & \tt{Sb} & 81.0& 1 & 73.7 & \tt{} & \tt{} \\
 \tt{NGC4236} & \tt{SBd} & 90.0 & 1 & 11.2 & \tt{} & \tt{} \\
 \tt{NGC4355} & \tt{SABa} & 68.0& 2 & 26.7 & \tt{} & \tt{} \\
 \tt{NGC4419} & \tt{Sa} & 84.5 & 2 & 6.0& \tt{} & \tt{} \\
 \tt{NGC4438} & \tt{Sa} & 73.2 & 1 & 25.4 & \tt{} & \tt{} \\
 \tt{NGC4527} & \tt{SABb} & 81.2 & 1 & 5.0& \tt{} & \tt{} \\
 \tt{NGC4772} & \tt{Sa} & 67.3 & 1 & 5.2 & \tt{} & \tt{} \\
 \tt{NGC4848} & \tt{Sc} & 73.9 & 1 & 29.0& \tt{} & \tt{} \\
 \tt{NGC4939} & \tt{Sbc} & 70.1 & 1 & 15.0& \tt{} & \tt{} \\
 \tt{NGC5394} & \tt{SBb} & 70.8 & 1 & 16.1 & \tt{} & \tt{} \\
 \tt{NGC5395} & \tt{SABb} & 66.1 & 1 & 16.1 & \tt{} & \tt{} \\
 \tt{NGC5674} & \tt{SABc} & 80.2 & 1 & 5.1 & \tt{} & \tt{} \\
 \tt{NGC6027C} & \tt{SBc} & 86.8 & 1 & 70.0& \tt{} & \tt{} \\
 \tt{NGC6503} & \tt{Sc} & 73.5 & 2 & 15.4 & \tt{} & \tt{} \\
 \tt{NGC6872} & \tt{SBb} & 72.5 & 2 & 76.3 & \tt{} & \tt{} \\
 \tt{NGC6925} & \tt{Sbc} & 84.1 & 1 & 10.0& \tt{} & \tt{} \\
 \tt{NGC7541} & \tt{SBc} & 74.8 & 1 & 39.5 & \tt{} & \tt{} \\
 \tt{NGC7591} & \tt{SBbc} & 66.9 & 1 & 5.0& \tt{} & \tt{} \\
 \tt{NGC7673} & \tt{Sc} & 68.2 & 1 & 59.4 & \tt{} & \tt{} \\
 \tt{NGC7753} & \tt{SABb} & 82.1 & 1 & 12.2 & \tt{} & \tt{} \\
 \tt{PGC001221} & \tt{SBc} & 73.3 & 1 & 30.0& \tt{} & \tt{} \\
 \tt{PGC019078} & \tt{E?} & 73.0& 1 & 14.8 & \tt{} & \tt{} \\
 \tt{PGC027508} & \tt{SBab} & 90.0 & 2 & 48.9 & \tt{} & \tt{} \\
 \tt{PGC038430} & \tt{Sd} & 75.8 & 1 & 15.0& \tt{} & \tt{} \\
 \tt{PGC046114} & \tt{Sbc} & 82.8 & 1 & 15.0& \tt{} & \tt{} \\
 \tt{PGC046133} & \tt{Sbc} & 80.5 & 1 & 15.0& \tt{} & \tt{} \\
 \tt{PGC086247} & \tt{Sbc} & 90.0 & 1 & 16.3 & \tt{} & \tt{} \\
 \tt{PGC100170} & \tt{SBbc} & 71.1 & 1 & 25.6 & \tt{} & \tt{} \\
 \tt{PGC2793298} & \tt{Sa} & 90.0 & 1 & 91.0& \tt{} & \tt{} \\
 \tt{UGC01934} & \tt{Sbc} & 90.0 & 1 & 9.0& \tt{} & \tt{} \\
 \tt{UGC02238} & \tt{Sm} & 68.9 & 1 & 15.1 & \tt{} & \tt{} \\
 \tt{UGC02626} & \tt{Sa} & 90.0 & 1 & 25.5 & \tt{} & \tt{} \\
 \tt{UGC03326} & \tt{Sc} & 90.0 & 1 & 80.1 & \tt{} & \tt{} \\
 \tt{UGC03995} & \tt{Sbc} & 66.4 & 1 & 11.0& \tt{} & \tt{} \\
 \tt{ESO121-006} & \tt{Sc} & 90.0 & \tt{} & \tt{} & 1 & 15.0\\
 \tt{ESO140-043} & \tt{SBb} & 72.8 & \tt{} & \tt{} & 2 & 49.4 \\
 \tt{ESO154-023} & \tt{SBm} & 90.0 & \tt{} & \tt{} & 1 & 18.7 \\
 \tt{ESO195-005} & \tt{Sa} & 90.0 & \tt{} & \tt{} & 35 & 581.3 \\
 \tt{ESO208-034} & \tt{SBab} & 75.4 & \tt{} & \tt{} & 2 & 29.2 \\
 \tt{ESO209-012} & \tt{Sa} & 90.0 & \tt{} & \tt{} & 1 & 20.4 \\
 \tt{ESO365-001} & \tt{Sc} & 90.0 & \tt{} & \tt{} & 1 & 14.9 \\
 \tt{ESO365-016} & \tt{SBab} & 66.4 & \tt{} & \tt{} & 1 & 26.3 \\
 \tt{ESO471-006} & \tt{SBm} & 90.0 & \tt{} & \tt{} & 1 & 19.9 \\
 \tt{ESO491-021} & \tt{SBab} & 79.0& \tt{} & \tt{} & 1 & 20.3 \\
 \tt{IC1504} & \tt{Sb} & 80.8 & \tt{} & \tt{} & 1 & 18.6 \\
 \tt{IC1537} & \tt{Sc} & 65.7 & \tt{} & \tt{} & 1 & 33.6 \\
 \tt{IC1959} & \tt{SBm} & 90.0 & \tt{} & \tt{} & 1 & 14.0\\
 \tt{IC4518A} & \tt{Sc} & 73.3 & \tt{} & \tt{} & 2 & 37.1 \\
 \tt{IC4518B} & \tt{Sc} & 90.0 & \tt{} & \tt{} & 2 & 37.1 \\
 \tt{IC5052} & \tt{SBcd} & 90.0 & \tt{} & \tt{} & 1 & 19.5 \\
 \tt{NGC0092} & \tt{Sa} & 69.9 & \tt{} & \tt{} & 1 & 48.0\\
 \tt{NGC0192} & \tt{SBa} & 76.2 & \tt{} & \tt{} & 1 & 48.9 \\
 \tt{NGC0675} & \tt{Sa} & 73.9 & \tt{} & \tt{} & 1 & 39.9 \\
 \tt{NGC0716} & \tt{Sa} & 76.2 & \tt{} & \tt{} & 1 & 25.6 \\
 \tt{NGC0784} & \tt{SBd} & 90.0 & \tt{} & \tt{} & 1 & 18.0\\
 \tt{NGC1134} & \tt{Sb} & 77.2 & \tt{} & \tt{} & 1 & 24.8 \\
 \tt{NGC1311} & \tt{SBm} & 90.0 & \tt{} & \tt{} & 1 & 14.8 \\
 \tt{NGC1320} & \tt{Sa} & 80.5 & \tt{} & \tt{} & 1 & 17.1 \\
 \tt{NGC1511} & \tt{Sab} & 73.7 & \tt{} & \tt{} & 1 & 44.9 \\
 \tt{NGC1512} & \tt{Sa} & 68.3 & \tt{} & \tt{} & 1 & 71.6 \\
 \tt{NGC2369} & \tt{Sa} & 90.0 & \tt{} & \tt{} & 2 & 47.5 \\
 \tt{NGC2613} & \tt{Sb} & 90.0 & \tt{} & \tt{} & 2 & 75.2 \\
 \tt{NGC3044} & \tt{SBc} & 90.0 & \tt{} & \tt{} & 2 & 38.0\\
 \tt{NGC3227} & \tt{SABa} & 68.3 & \tt{} & \tt{} & 4 & 162.6 \\
 \tt{NGC3281} & \tt{Sab} & 71.7 & \tt{} & \tt{} & 1 & 23.7 \\
 \tt{NGC3735} & \tt{Sc} & 83.5 & \tt{} & \tt{} & 2 & 21.9 \\
 \tt{NGC3746} & \tt{Sab} & 66.6 & \tt{} & \tt{} & 2 & 61.0\\
 \tt{NGC3753} & \tt{Sab} & 90.0 & \tt{} & \tt{} & 2 & 61.0\\
 \tt{NGC3786} & \tt{SABa} & 65.2 & \tt{} & \tt{} & 1 & 29.5 \\
 \tt{NGC3788} & \tt{SABa} & 86.1 & \tt{} & \tt{} & 1 & 29.5 \\
 \tt{NGC3976} & \tt{SABb} & 81.6 & \tt{} & \tt{} & 1 & 14.4 \\
 \tt{NGC4157} & \tt{SABb} & 90.0 & \tt{} & \tt{} & 1 & 62.7 \\
 \tt{NGC4173} & \tt{SBcd} & 90.0 & \tt{} & \tt{} & 1 & 12.9 \\
 \tt{NGC4235} & \tt{Sa} & 90.0 & \tt{} & \tt{} & 1 & 13.1 \\
 \tt{NGC4302} & \tt{Sc} & 90.0 & \tt{} & \tt{} & 2 & 100.8 \\
 \tt{NGC4319} & \tt{SBab} & 72.5 & \tt{} & \tt{} & 7 & 246.4 \\
 \tt{NGC4330} & \tt{Sc} & 78.8 & \tt{} & \tt{} & 1 & 31.4 \\
 \tt{NGC4437} & \tt{Sc} & 90.0 & \tt{} & \tt{} & 1 & 114.3 \\
 \tt{NGC4536} & \tt{SABb} & 73.0& \tt{} & \tt{} & 1 & 34.2 \\
 \tt{NGC4605} & \tt{SBc} & 70.0& \tt{} & \tt{} & 1 & 8.0\\
 \tt{NGC4634} & \tt{SBc} & 80.5 & \tt{} & \tt{} & 3 & 135.9 \\
 \tt{NGC4686} & \tt{Sa} & 82.7 & \tt{} & \tt{} & 2 & 32.6 \\
 \tt{NGC4700} & \tt{SBc} & 90.0 & \tt{} & \tt{} & 1 & 87.3 \\
 \tt{NGC4845} & \tt{Sab} & 90.0 & \tt{} & \tt{} & 2 & 29.7 \\
 \tt{NGC5073} & \tt{SBc} & 90.0 & \tt{} & \tt{} & 1 & 53.6 \\
 \tt{NGC5356} & \tt{SABb} & 90.0 & \tt{} & \tt{} & 2 & 38.7 \\
 \tt{NGC5899} & \tt{Sc} & 68.9 & \tt{} & \tt{} & 2 & 29.6 \\
 \tt{NGC6045} & \tt{SBc} & 85.3 & \tt{} & \tt{} & 3 & 75.2 \\
 \tt{NGC6323} & \tt{Sab} & 68.3 & \tt{} & \tt{} & 2 & 25.7 \\
 \tt{NGC6810} & \tt{Sab} & 90.0 & \tt{} & \tt{} & 1 & 48.7 \\
 \tt{NGC6814} & \tt{SABb} & 85.6 & \tt{} & \tt{} & 2 & 36.1 \\
 \tt{NGC6926} & \tt{Sc} & 78.1 & \tt{} & \tt{} & 1 & 11.8 \\
 \tt{NGC7314} & \tt{SABb} & 69.8 & \tt{} & \tt{} & 2 & 55.2 \\
 \tt{PGC006966} & \tt{Sc} & 90.0 & \tt{} & \tt{} & 1 & 14.9 \\
 \tt{PGC012596} & \tt{Sb} & 71.6 & \tt{} & \tt{} & 1 & 17.9 \\
 \tt{PGC013944} & \tt{SBab} & 70.8 & \tt{} & \tt{} & 1 & 18.9 \\
 \tt{PGC014121} & \tt{S?} & 77.6 & \tt{} & \tt{} & 1 & 38.8 \\
 \tt{PGC023515} & \tt{Sa} & 65.7 & \tt{} & \tt{} & 1 & 11.9 \\
 \tt{PGC026440} & \tt{SABc} & 66.6 & \tt{} & \tt{} & 1 & 16.0\\
 \tt{PGC037282} & \tt{Scd} & 74.4 & \tt{} & \tt{} & 1 & 19.6 \\
 \tt{PGC044532} & \tt{Sm} & 90.0 & \tt{} & \tt{} & 3 & 64.9 \\
 \tt{PGC053471} & \tt{Sc} & 90.0 & \tt{} & \tt{} & 2 & 37.1 \\
 \tt{PGC061664} & \tt{Sb} & 86.0& \tt{} & \tt{} & 2 & 31.3 \\
 \tt{PGC063176} & \tt{SABa} & 81.4 & \tt{} & \tt{} & 8 & 132.2 \\
 \tt{PGC064775} & \tt{Sd} & 90.0 & \tt{} & \tt{} & 1 & 14.2 \\
 \tt{PGC065349} & \tt{SABa} & 65.6 & \tt{} & \tt{} & 2 & 20.9 \\
 \tt{PGC066146} & \tt{Sc} & 68.7 & \tt{} & \tt{} & 2 & 32.1 \\
 \tt{PGC074302} & \tt{SABa} & 90.0 & \tt{} & \tt{} & 1 & 16.1 \\
 \tt{PGC402573} & \tt{Sbc} & 67.1 & \tt{} & \tt{} & 2 & 31.5 \\
 \tt{UGC00717} & \tt{Sb} & 65.8 & \tt{} & \tt{} & 1 & 15.0\\
 \tt{UGC00987} & \tt{Sa} & 90.0 & \tt{} & \tt{} & 1 & 22.7 \\
 \tt{UGC08515} & \tt{Sab} & 90.0 & \tt{} & \tt{} & 1 & 8.1 \\
 \tt{UGC09944} & \tt{Sbc} & 79.6 & \tt{} & \tt{} & 3 & 41.9 \\  \hline
 {\tt Sum:} & & & & 7078.3 & & 8660.7\\
\end{longtable}
\end{center}
\end{small}

\bibliography{refs}
\bibliographystyle{JHEP}

\end{document}